\documentclass[aps,pra,reprint,superscriptaddress]{revtex4-1}

\usepackage{lmodern}
\usepackage[T1]{fontenc}
\usepackage[english,activeacute]{babel}
\usepackage{mathtools}
\usepackage{color} 

\usepackage{amssymb} 
\usepackage{dsfont} 

\usepackage{pbsi}

\usepackage[colorlinks,linkcolor=blue,citecolor=blue,urlcolor=blue]{hyperref}

\addto\extrasenglish{}
\addto\extrasenglish{}

\begin{document}


\newcommand{\dd}{\mbox{d}}
\newcommand{\ii}{\mbox{i}}
\newcommand{\ddd}{\mbox{\scriptsize{d}}}
\newcommand{\iii}{\mbox{\scriptsize{i}}}

\newcommand{\QE}{\mbox{\tiny{QE}}}
\newcommand{\LL}{\mbox{\tiny{L}}}
\newcommand{\C}{\mbox{\tiny{C}}}

\newcommand{\LP}{\mbox{\tiny{LP}}}
\newcommand{\UP}{\mbox{\tiny{UP}}}

\newcommand{\JC}{\mbox{\tiny{JC}}}

\newcommand{\g}{\mbox{g}}
\newcommand{\e}{\mbox{e}}
\newcommand{\gmini}{\mbox{\scriptsize{g}}}
\newcommand{\emini}{\mbox{\scriptsize{e}}}
\newcommand{\eg}{\mbox{\scriptsize{eg}}}

\newcommand{\eff}{\mbox{\scriptsize{eff}}}

\newcommand{\intt}{\mbox{\scriptsize{int}}}

\newcommand{\Tr}{\mbox{Tr}}

\newcommand{\HH}{\mbox{\tiny{H}}}
\newcommand{\VV}{\mbox{\tiny{V}}}

\newcommand{\II}{\mbox{\tiny{I}}}

\newcommand{\rr}{\mbox{\tiny{r}}} 
\newcommand{\nr}{\mbox{\tiny{nr}}} 

\newcommand{\dtQE}{\tilde{\Delta}_{\QE}} %
\newcommand{\dtC}{\tilde{\Delta}_{\C}} %

\newcommand{\Ed}{\boldsymbol{E}_{\rm D}}

\newcommand{\Gz}{G^{(2)}(0)}
\newcommand{\Gtau}{G^{(2)}(\tau)} %
\newcommand{\gz}{g^{(2)}(0)} %
\newcommand{\gtau}{g^{(2)}(\tau)} %

\title{Photon Statistics in Collective Strong Coupling: \\ Nano- and Microcavities}

\author{R. S\'aez-Bl\'azquez}
\email{rocio.saezblazquez@uam.es}
\affiliation{Departamento de F\'isica Te\'orica de la Materia Condensada and Condensed Matter Physics Center (IFIMAC), Universidad Aut\'onoma de
    Madrid, E-28049 Madrid, Spain}
\author{J. Feist}
\affiliation{Departamento de F\'isica Te\'orica de la Materia Condensada and Condensed Matter Physics Center (IFIMAC), Universidad Aut\'onoma de
    Madrid, E-28049 Madrid, Spain}
\author{F. J. Garc\'ia-Vidal}
\affiliation{Departamento de F\'isica Te\'orica de la Materia Condensada and Condensed Matter Physics Center (IFIMAC), Universidad Aut\'onoma de
    Madrid, E-28049 Madrid, Spain}
\affiliation{Donostia International Physics Center
    (DIPC), E-20018 Donostia/San Sebasti\'an, Spain}
\author{A. I. Fern\'andez-Dom\'inguez}
\affiliation{Departamento de F\'isica Te\'orica de la Materia Condensada and Condensed Matter Physics Center (IFIMAC), Universidad Aut\'onoma de
    Madrid, E-28049 Madrid, Spain}

\begin{abstract}
There exists a growing interest in the properties of the light
generated by hybrid systems involving a mesoscopic number of
emitters as a means of providing macroscopic quantum light
sources. In this work, the quantum correlations of the light
emitted by a collection of emitters coupled to a generic optical
cavity are studied theoretically using an effective Hamiltonian
approach. Starting from the single-emitter level, we analyse the
persistence of photon antibunching as the ensemble size increases.
Not only is the photon blockade effect identifiable, but
photon antibunching originated from destructive interference
processes---the so-called unconventional antibunching---is also
present. We study the dependence of these two types of negative
correlations on the spectral detuning between cavity and emitters,
as well as its evolution as the time delay between photon
detections increases. Throughout this work, the performance of
plasmonic nanocavities and dielectric microcavities is compared: despite
the distinct energy scales and the differences introduced by their
respectively open and closed character, the bunching and antibunching
phenomenology presents remarkable similarities in both types of cavities.
\end{abstract}

\maketitle

\section{Introduction}

The field of cavity quantum electrodynamics studies phenomena
arising from the interaction between matter and light---the latter
in the form of electromagnetic modes confined in a cavity---in the
regime in which the quantum nature of the light is
unveiled~\cite{MicrocavitiesBook}. The matter component commonly
involves two-level systems, which describe a wide variety of
emitters (from real spins to quantum dots, atoms, molecules,
NV-centers, or qubits, among others) as long as the state of the
system can be properly represented by a two-dimensional Hilbert
space. The most simple case of a single two-level system coupled
to a quantized cavity mode is described by the  Jaynes-Cummings
model~\cite{JaynesCummings1963}, later extended to an arbitrary
number of emitters in the so-called  Tavis-Cummings
model~\cite{TavisCummings1968}. These Hamiltonians, together with
specific procedures to introduce the effect of
losses~\cite{OpenBook}, provide a theoretical description for a
broad variety of configurations.

The interaction of light with a collection of quantum emitters has
been intensively studied considering an extensive range of
systems, since it is of broad interest in research areas ranging
from lasing~\cite{Lasing1, Lasing2} and
superradiance~\cite{Haroche1982, Hommel2007} to non-classical
light generation~\cite{VuckovicNat2008},
entanglement~\cite{Haroche2001}, or quantum information
processing~\cite{Zoller1995, Imamoglu1999}. In fact, there are
several optical cavity configurations capable of supporting an
electromagnetic mode through semiconductor, metal, or dielectric
structures, suiting thereby the specific requirements. From high-quality optical microcavities~\cite{MicrocavitiesBook,
    Vahala2003} (consisting in different planar
configurations~\cite{pillarMicrocavity,braggMicrocavity},
whispering-galleries~\cite{whisper1, whisper2}, or
photonic-crystal cavities~\cite{microPhotCrystal1,
    microPhotCrystal2,nanoPhotCrystal}, to name a few), the aim to
reach higher light-matter couplings---especially on the route
towards room-temperature devices---has led to the reduction of
the effective volume even below the diffraction limit of classical
optics. This has given rise to nanocavities built on the basis of
plasmonic nanostructures. The price to pay for the large confinement
is that these structures suffer from large dissipative losses,
reducing in turn the quality factor of the cavity.

The balance between coupling strength and losses leads to the
well-known distinction between the weak and strong coupling
regimes. Within the weak coupling regime, the principal feature
consists in the enhancement of the spontaneous emission owing to
the coupling of the emitters to the resonant cavity, known as
the Purcell effect~\cite{Purcell1946}. On the contrary, if the
matter-field interaction becomes larger than the emitter and
cavity relaxation rates, the system may enter the strong coupling
regime, in which the genuine eigenstates turn out to be a quantum
mixture of matter and light. These are referred to as dressed
states or polaritons, arising due to the rapid exchange of energy
between cavity and emitters. The regime of strong coupling  has
been demonstrated experimentally for systems involving a large
amount of emitters for
microcavities~\cite{Arakawa1992,Weisbuch1994} as well as for
plasmonic nanocavities~\cite{Bellesa2004,Ebbesen2011}. Reaching
this regime by involving just a single emitter becomes more
challenging, since it requires a cavity with a higher quality
factor or stronger field confinement. While experimental realizations of single-emitter strong
coupling have been reported in microcavities~\cite{Kimble2003,
    Forchel2004} and photonic crystals~\cite{Yoshie2004} in the past,
only recently has the strong coupling regime been reached in
plasmonic nanocavities with a single molecule~\cite{Baumberg2016,
    Santhosh2016}. Note that by varying the number of emitters, the
system can be naturally tuned from one regime to the other---in
these hybrid systems, the participation of a collection of $N$
quantum emitters makes the interaction stronger through the
appearance of the characteristic $\sqrt{N}$ factor in the
effective collective coupling strength~\cite{Dicke1954}.

Whereas coupling a large amount of emitters to a nanocavity mode
is not difficult from the experimental perspective, addressing the
complete problem theoretically presents a significant challenge. There have
been some theoretical advances in the treatment of plasmonic
strong coupling for large ensembles of emitters in the vanishing
population regime~\cite{Delga2014}. There, the fermionic character
of quantum emitters can be neglected by modelling them as bosonic
harmonic oscillators and, thus, by disregarding excitonic
nonlinearities and any saturation effects emerging in the photon
population dynamics for the system. One step beyond this
completely bosonic description was also carried out through the
so-called Holstein-Primakoff approach~\cite{GonzalezTudela2013, DeLiberato2017},
which is equivalent to introducing a macroscopic third-order
susceptibility for the medium embedding the emitter
ensemble~\cite{Alpeggiani2016}. Very recently, we proposed a
theoretical framework~\cite{Rocio2017} able to describe
plasmon-exciton strong coupling for a mesoscopic number of quantum
emitters which fully accounted for their fermionic character. It
is precisely the inherent quantum nonlinear character of emitters
that forms the basis of many interesting phenomena when coupling
light and matter, such as the well-known photon
antibunching~\cite{KimbleMandel1977}. In general, intensity
correlations of the emitted light are related to the probability
of detecting coincident photons, so that it is used as a
quantifier of multiple-photon events (the supression of which is
an important requirement for single-photon emission). Importantly,
although this magnitude has been extensively used to discriminate
single-photon sources, some recent papers point out that this is
not by itself a reliable indicator~\cite{Laussy}.

Non-classical light generation is extremely important in the
fields of quantum cryptography~\cite{quantumCryptography}, quantum
sensing~\cite{quantumSensing}, quantum
metrology~\cite{quantumMetrology}, or quantum
communication~\cite{Kimble2008}, among other emerging photonic
quantum technologies. The production of single photons on demand
has focused great efforts, and different methods have been used
for its extraction~\cite{SinglePhotons2005}, such as faint laser
pulses~\cite{faintPulses}, single-emitter
systems~\cite{singleAtom,singleMolecule,diamond}, non-linear
crystals~\cite{nanocrystal} or parametric
downconversion~\cite{parametricDownconversion}. All these exploit
the inherent non-linearity of the photonic system. When using a
large number of emitters, their collective response becomes
approximately bosonic and non-classical light is supposed not to
be generated. Nevertheless, it has been shown that when
considering a mesoscopic number of emitters, these non-linearities
can be preserved and antibunched light may still be
produced~\cite{Kimble1991, Carmichael1991, Foster2000}. Much
effort has been devoted to the analysis of the evolution of the
quantum statistical properties of the light emitted by hybrid
systems for increasing number of emitters, assessing the
possibility of generating non-classical light with mesoscopic
ensembles. Some studies involve just a few quantum
emitters~\cite{Auffeves2011, Radulaski2017}, and others consider a
huge amount under particular
simplifications~\cite{MeiserHolland2010a,Richter2015,Grangier2014}.
From a technical perspective, there exist brute-force approaches
based on Monte-Carlo techniques~\cite{TemnovWoggon2005,
TemnovWoggon2009}, and developments towards the efficient
treatment of large ensembles~\cite{Richter2016} thanks to the use
of symmetries.

In this article, we study the coherence properties of the light
radiated by a collection of $N$ identical quantum emitters placed
inside a generic optical cavity when the system is coherently
pumped by a laser. Two paradigmatic configurations are explored,
namely, a collection of quantum dots coupled to a dielectric
microcavity, and an ensemble of organic molecules within a
plasmonic nanocavity (note that the pumping and the emission
differ according to the open or closed character, respectively, of
these systems). Whereas the former have been extensively
investigated within the field of semiconductor quantum
electrodynamics over the last two decades, the exploration of the
latter for quantum optical purposes is still in a very early
stage. To our knowledge, this article provides the first
comparative study between both physical platforms for
non-classical light generation. In order to perform a meaningful
comparison between both physical systems, we treat them in the
same footing. In turn, this means that aspects that may be
relevant in specific experimental implementations of both
configurations, such as spatial and spectral inhomogeneities or
emitter-emitter interactions, are neglected. The statistics of the
emitted light are analysed for these two cases determining the
parameter ranges in which photon bunching and antibunching appear.
In particular, the focus is on the two effects that can lead to
single-photon emission: the well-known {\it photon blockade
    effect} and the so-called {\it unconventional antibunching}, which
originates from interference effects and is thus here referred to as
{\it interference-induced correlations}. On the basis of the
quantum master equation for the extended Tavis-Cummings model,
analytical and numerical computations are performed by using an
effective Hamiltonian approach.

Our theoretical framework is described in
\autoref{sec:TheoreticalFramework}, beginning with the
introduction of the two systems under consideration and their
modelling (\autoref{subsec:TF_System}). The procedure to determine
the steady state of the system (\autoref{subsec:TF_SteadyState})
and compute the correlation functions
(\autoref{subsec:TF_CorrelationFunctions}) is described next. In
\autoref{sec:Results}, we present a comprehensive analysis of
photon statistics for realistic plasmonic nanocavities and
dielectric microcavities. First, the intensity and second-order
correlation function of the emitted light are explored for various
ensemble sizes (\autoref{subsec:R_IntensityCorrelation}), and
analytical expressions for these magnitudes for both cavity
configurations are provided. Then, the two different mechanisms
leading to sub-Poissonian light are studied in detail
(\autoref{subsec:R_TwoMechanisms}), followed by the analysis of
the effect of spectral detuning between cavity and emitters on the
photon correlations (\autoref{subsec:R_Detuning}). The evolution
of the second-order correlation function at time delays different
from zero is also investigated (\autoref{subsec:R_Tau}). Finally,
the conclusions of the work are presented in
\autoref{sec:Conclusions}.

\section{Theoretical framework} \label{sec:TheoreticalFramework}

\subsection{System} \label{subsec:TF_System}
\begin{figure}[ht]
    \centering
    \includegraphics[width=\linewidth]{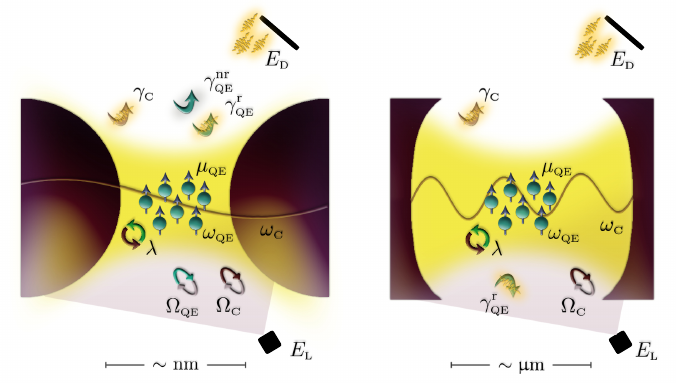}
    \caption{Scheme of the systems, composed of a collection of quantum emitters coupled to an electromagnetic mode supported by either a nano- (left) or a micro- (right) cavity. The set of parameters characterising the system are sketched.}
    \label{fig:modelScheme}
\end{figure}

The system under study consists of $N$ quantum emitters---modelled
as simple two-level systems with a ground $| \g_n \rangle$ and an
excited state $| \e_n \rangle$---located in an optical cavity.
Every emitter is coupled to a quantized single cavity mode through
the electric-dipole interaction, and they are considered not to
interact among them apart from through the cavity mode (note that
emitter-emitter interactions become significant only in dense
ensembles~\cite{Rocio2017}). A laser field coherently pumps the
system, and the emitted light is collected in a detector located
in the far-field. An illustration of the system is depicted in
\autoref{fig:modelScheme}, where the two cases of a nano- (left)
and a micro- (right) cavity are distinguished. Both the pumping
and the emission varies according to the corresponding open or
closed character: whereas for nanocavities the entire system is
pumped and the radiation from both the quantum emitters and the
cavity is observed (open configuration), in microcavities the
coupling to the outside mode is mediated by the mirrors, such that
only the cavity mode is pumped and only its emission is received
at the detector (closed configuration). Apart from this
fundamental distinction, the size and characteristic losses are
also differentiating features between these two types of cavities.
By nanocavities, we are referring to plasmonic
cavities~\cite{SavastaACS2010,Sandoghdar2013,
Shegai2015,Peyskens2017}, where the spatial dimensions are reduced
to the nanometre scale and cavity losses are substantial ($\sim
0.1$~eV)~\cite{Tame2013}. In contrast, by microcavities we refer
here to photonic crystals~\cite{Noda2003,Englund2007} and other
semiconductor structures~\cite{Jewell1989,Forchel2004} with sizes
of the order of micrometres and whose absorption is much lower
($\sim 0.1$~meV)~\cite{Vahala2003}.

The Hamiltonian of the total system involves the excitation of both the
cavity mode (with transition frequency $\omega_{\C}$ and bosonic
creation and annihilation operators $a^\dagger$ and $a$) and the
collection of $N$ quantum emitters (with transition frequency
$\omega_n$ and creation and annihilation operators
$\sigma_n^\dagger = |\e_n \rangle \langle \g_n |$ and $\sigma_n =
|\g_n \rangle \langle \e_n |$ for the $n$-th emitter which, in
turn, define the operator $\sigma_n^z = [\sigma_n^\dagger,
\sigma_n]/ 2 $), as well as the coherent pumping of a laser of frequency $\omega_{\LL}$. In the Schr\"odinger picture, it reads
(from now on, $\hbar = 1$):
\begin{equation}
\begin{split}
H  = \ & \omega_{\C} \ a ^\dagger  a +  \sum_{n=1}^N  \omega_{n}  \sigma_n^z
+   \sum_{n=1}^N  \lambda_n (a^\dagger  \sigma_n + a \sigma_n^\dagger) \\
& +   \Omega_{\C} ( a ^\dagger \ e^{- \iii \omega_{\LL} t}   +   a \ e^{ \iii \omega_{\LL} t}) \\
& +  \sum_{n=1}^N  \Omega_{n} ( \sigma_n^\dagger \ e^{- \iii \omega_{\LL} t}   + \sigma_n  \ e^{ \iii \omega_{\LL} t}) \ ,
\end{split}
\label{eq:HamiltonianSP}
\end{equation}
where $\lambda_n$ is the coupling between the $n$-th quantum
emitter and the cavity mode, and $\Omega_{\C}$ and $\Omega_n$ are
the laser pumping to the cavity mode and the $n$-th quantum
emitter respectively. If we define the effective dipole moments
associated with the cavity,  $\boldsymbol{\mu}_{\C}$, and with the
$n$-th emitter, $\boldsymbol{\mu}_{n}$, these pumping frequencies
can be expressed in terms of the laser field intensity
$\boldsymbol{ E}_{\LL}$ as $\Omega_{\C} = \boldsymbol{ E}_{\LL}
\cdot \bf {\boldsymbol{\mu}_{\C}}$ and $\Omega_{n} =
\boldsymbol{E}_{\LL} \cdot \boldsymbol{\mu}_{n}$. In the
description of a microcavity, only the cavity is pumped---hence
we set $\Omega_n = 0$ for all $n$. Note also that the
{\it rotating wave approximation}~\cite{WallsMilburnBook} has been
introduced in \autoref{eq:HamiltonianSP}, which implies that the
emitter-cavity coupling is low enough to disregard the fast-rotating terms.

Through the coherent excitation, the Hamiltonian
(\ref{eq:HamiltonianSP}) acquires an explicit dependence on time,
which can be easily removed by transforming it into a rotating
frame. By means of the unitary operator
$
\mathcal{U}_0(t) = \mbox{exp}[{- \ii A t}] \ ,
$
with
$
A =   \omega_{\LL} a^\dagger a + \sum_{n=1}^N  \omega_{\LL} \sigma_n^z \ ,
$
the Hamiltonian in the transformed frame,
$\tilde H = \mathcal{U}_0^\dagger H \mathcal{U}_0 - A$,
becomes:
\begin{equation}
\begin{split}
\tilde  H  = \ &  \Delta_{\C} \ a ^\dagger  a +  \sum_{n=1}^N  \Delta_{n}  \sigma_n^z
+   \sum_{n=1}^N  \lambda_n (a^\dagger  \sigma_n + a \sigma_n^\dagger) \\
& +   \Omega_{\C} ( a ^\dagger     +   a )
+  \sum_{n=1}^N  \Omega_{n} ( \sigma_n^\dagger + \sigma_n)  \ ,
\end{split}
\label{eq:HamiltonianIP}
\end{equation}
where the detunings from the laser corresponding to the cavity
mode, $ \Delta_{\C} = \omega_{\C} - \omega_{\LL}$, and the $n$-th
quantum emitter, $ \Delta_{n} = \omega_n - \omega_{\LL}$, have
been defined.

When all quantum emitters have the same transition frequency
$\omega_{\QE}$ (consequently, the same detuning $\Delta_{\QE} =
\omega_{\QE} - \omega_{\LL}$) and equal couplings to the cavity,
$\lambda$, and to the laser field, $\Omega_{\QE} =
\boldsymbol{E}_{\LL} \cdot \boldsymbol{\mu}_{\QE}$, the above
expression for the Hamiltonian (\ref{eq:HamiltonianIP}) reduces
to:
\begin{equation}
\begin{split}
\tilde  H  = \ & \Delta_{\C} \ a ^\dagger  a +    \Delta_{\QE}  S^z
+   \lambda (a^\dagger  S^- + a S^+) \\
&
+   \Omega_{\C} ( a ^\dagger     +   a )
+   \Omega_{\QE} ( S^+ + S^-)  \ ,
\end{split}
\label{eq:HamiltonianIPbright}
\end{equation}
where the bright mode creation and annihilation operators, $S^+ =
\sum_{n=1}^N \sigma_n^\dagger $ and $S^- = (S^+)^\dagger$,
together with the operator $S^z = \sum_{n=1}^N \sigma_n^z$, have
been introduced for this set of $N$ emitters. These collective operators
behave like standard spin-$N/2$ operators.
Note that the coupling $\lambda$ is just the interaction between the emitter
dipole moment and the near field of the cavity, that is, $\lambda
= \boldsymbol{E}_{\C} \cdot \boldsymbol{\mu}_{\QE}$. Some tests
beyond these assumptions, treating the emitters and their
respective couplings individually, can be found in
Ref.~\cite{Rocio2017}.

\subsection{Steady state of the system} \label{subsec:TF_SteadyState}

To study the properties of the emitted light, we first need to
determine the steady-state of the system. The time-evolution of
the density matrix $\rho$ describing the system is governed by the
following master equation:
\begin{equation}
\frac{\dd}{\dd t} \rho = - \ii \ [\tilde H, \rho]
+ \frac{\gamma_{\C}}{2}\mathcal{L}_a[\rho] +\frac{\gamma_{\QE}^{\rr}}{2} \mathcal{L}_{S^-}[\rho] + \sum_{n=1}^N \frac{\gamma_{\QE}^{\nr}}{2} \mathcal{L}_{\sigma_n}[\rho]\ ,
\label{eq:MasterEquation}
\end{equation}
where the Lindblad terms, given by $\mathcal{L}_\mathcal{O} = 2
\mathcal{O} \rho  \mathcal{O}^\dagger - \mathcal{O}^\dagger
\mathcal{O} \rho - \rho \mathcal{O}^\dagger \mathcal{O}$ for an
arbitrary operator $\mathcal{O}$, account for the losses arising
from both cavity and quantum emitters (with decay rates
$\gamma_{\C}$ and $\gamma_{\QE}$ respectively). The superscripts
stand for radiative (r)  and non-radiative (nr) damping. Note that
while all cavity losses are included in $\mathcal{L}_a$, two
different terms have to be added for the quantum emitters:
radiative losses are described through the bright mode operator
$S^-$ (corresponding to the assumption that the emitters are at
sub-wavelength distances and thus radiate like a collective dipole),
but non-radiative losses are assigned to single emitters.
Therefore, it is the single-atom operator $\sigma_n$ that is
involved in the latter case.

In the regime of sufficiently low driving intensity, the contribution
of the so-called {\it refilling} or {\it feeding} terms
$\mathcal{O} \rho \mathcal{O}^\dagger$ appearing in the Lindblad
superoperators remains negligible~\cite{Rocio2017}. When they are
removed, the Lindblad master equation (\ref{eq:MasterEquation}) becomes
equivalent to the Schr\"odinger equation $\dd |\psi \rangle
/ \dd t = - \ii H_{\rm eff} | \psi \rangle$ with an effective
Hamiltonian:
\begin{equation}
H_{\rm eff} = \tilde{H} - \ii \frac{\gamma_{\C}}{2} a^\dagger a
- \ii \frac{\gamma_{\QE}^{\rr}}{2} S^+ S^-
- \ii \frac{\gamma_{\QE}^{\nr}}{2} S^z   \ ,
\label{eq:EffectiveHamiltonian}
\end{equation}
where $\tilde H$ is given by \autoref{eq:HamiltonianIPbright}.
In this effective Hamiltonian, only bright mode operators appear.
Within this approach, the dark states
of the ensemble, superpositions of the quantum emitter excitations
that do not couple to the cavity or the external light, can thus be
disregarded without further approximation. This corresponds to a
crucial reduction in numerical effort when
considering a large number of emitters $N$: instead of having to
consider $N$ and $N(N-1)$ states in the one- and two-excitation
manifold respectively, only one singly and one doubly excited bright
state, $\sum_{n=1}^{N} \sigma_n^\dagger | 0 \rangle/ \sqrt{N}$
and $\sum_{n, m =1}^{N} \sigma_n^\dagger \sigma_m^\dagger
| 0 \rangle / \sqrt{N (N-1)}$ (with $n \neq m$), play a role.

In the low pumping regime, we can perturbatively solve the Schr\"odinger
equation $H_{\eff} |\psi \rangle = 0$ (equivalent to solving for the
steady state solution of the master equation $\dd \rho / \dd
t = 0$). Considering the incident laser amplitude $E_{\LL}$ as the small
parameter, the effective Hamiltonian
(\ref{eq:EffectiveHamiltonian}) can be split as $H_{\eff} = H_0 +
E_{\LL} V$, where the second term is the driving:
\begin{equation}
E_{\LL} V = \Omega_{\C}(a^\dagger + a) + \Omega_{\QE} (S^+ + S^-) \ .
\end{equation}
The steady state $| \psi \rangle$ can also be expanded in a power
series of $E_{\LL}$, $ | \psi \rangle = \sum_{k = 0} E_{\LL}^k |
\psi_k \rangle $. Substituting these expansions into the
equation $H_{\eff} |\psi \rangle= 0$ and grouping terms for each
power of $E_{\LL}$ results in a set of linear equations. The
zeroth-order equation leads simply to $|\psi_0 \rangle = |0
\rangle$ (that is, the ground state, which represents no
excitations in the system) whereas the $k$th-order equation turns
out to be $H_0 | \psi_k \rangle + V| \psi_{k-1} \rangle = 0$.
These equations can be successively solved so that the steady
state is finally obtained from this perturbative approach.

\subsection{First- and second-order correlation functions}  \label{subsec:TF_CorrelationFunctions}

Once the steady state is known, the correlation properties of the
emitted light can be calculated. The negative-frequency part of
the scattered far-field operator at the detector,
$\boldsymbol{E}_{\rm D}^-$, depends on the type of cavity we
consider: while for nanocavities the radiation from both cavity
and quantum emitters is taken into account, $\boldsymbol{E}_{\rm
    D}^- \propto \boldsymbol{\mu}_{\C} a^\dagger +
\boldsymbol{\mu}_{\QE} S^+$, for microcavities just the emission
coming from the cavity is detected, $\boldsymbol{E}_{\rm D}^-
\propto \boldsymbol{\mu}_{\C} a^\dagger$. This reflects the
open/closed character of each type of cavity. Note that the
differences between the electromagnetic Green's function
describing the emission from the cavity and the various emitters
in nanocavities can be neglected due to their deeply subwavelength
dimensions.  The light intensity at a given point $I ({\boldsymbol
    r}, t)$ is defined in terms of the electric field operator as:
\begin{equation}
I ({\boldsymbol r}, t) = \langle \Ed^- ({\boldsymbol r}, t) \Ed^+ ({\boldsymbol r}, t) \rangle
\end{equation}
and the two-time second-order correlator $G^{(2)}$ and its
normalized version $g^{(2)}$ are given by~\cite{LoudonBook}:
\begin{equation}
\begin{split}
G^{(2)} & ({\boldsymbol r_1}, t_1; {\boldsymbol r_2}, t_2)  = \\
 & \langle \Ed^- ({\boldsymbol r_1}, t_1)   \Ed^- ({\boldsymbol r_2}, t_2) \Ed^+ ({\boldsymbol r_2}, t_2) \Ed^+ ({\boldsymbol r_1}, t_1) \rangle
\\
g^{(2)} & ({\boldsymbol r_1}, t_1; {\boldsymbol r_2}, t_2)  = \frac{G^{(2)} ({\boldsymbol r_1}, t_1; {\boldsymbol r_
        2}, t_2)}{I ({\boldsymbol r_1}, t_1)    I ({\boldsymbol r_2}, t_2) } \ ,
\end{split}
\end{equation}
where $\langle \cdot \rangle$ denotes time average. The latter is
related to the (conditional/joint) probability of detecting a
photon in the detector placed at $r_2$ at time $t_2$ once a photon
has reached the detector placed at $r_1$ at time $t_1$.
Considering a fixed position, it can be rewritten in terms of the
time delay $\tau = t_2 - t_1$ as:
\begin{equation}
g^{(2)} (\tau)  = \frac{\langle \Ed^- (t)   \Ed^- (t + \tau) \Ed^+ (t + \tau) \Ed^+ (t) \rangle}
{I (t)    I (t + \tau) } \ ,
\end{equation}
which does not depend on time $t$ in the steady state. Note that
for $\tau = 0$, it yields the probability of detecting two
coincident photons. When $\gtau < \gz$, registering two photon
counts with a delay $\tau$ is less likely than the observation of
two simultaneous photons. This is known as {\it photon bunching},
since photons tend to be distributed close together, in ``bunches'',
instead of being located further apart. The opposite situation is
given when $\gtau > \gz$, known as {\it photon antibunching}, where
photons tend to arrive at different times. For a coherent source
of light, $\gtau = 1$ for all $\tau$, which means that photons
arrive independently from one another at the detector (note that
$\gtau \rightarrow 1$ when $\tau \rightarrow \infty$ for any light
source), leading to a Poissonian distribution of arrival times.
The statistics of the
light is then said to be {\it super-Poissonian} ($\gz
> 1$) or {\it sub-Poissonian} ($\gz<1$) if the coincidence of two
photons at the detector is, respectively, more or less likely than
that for a coherent light source (random case). The concepts of
antibunching and sub-Poissonian statistics are often not distinguished in
the literature since they usually occur together. Nevertheless, they
are not equivalent concepts but reflect distinct effects; indeed,
sub-Poissonian statistics can take place together with
bunching~\cite{ZouMandel1990}. Since the evolution of $\gtau$ with
time delay is explored in this article, we distinguish them
rigorously throughout the text to avoid misunderstandings. All the
same, both antibunching and sub-Poissonian statistics are
phenomena related to non-classical light, as the conditions
defining them cannot be fulfilled for classical
fields~\cite{WallsMilburnBookSTATISTICS}. Therefore, they offer a
means to measure the classicality/quantumness of light.

The first- and second-order correlation functions can be evaluated
by considering the perturbative solution for the steady state of
the system described above. The scattering intensity $I$ and the
normalized zero-delay second-order correlation function $\gz$ in
the steady state are thus computed as:
\begin{equation}
\begin{split}
I & = \langle \psi_1 | \Ed^- \Ed^+  | \psi_1 \rangle   \\
g^{(2)} (0) & = \langle \psi_2 | \Ed ^- \Ed^- \Ed^+ \Ed^+ | \psi_2 \rangle
/ I ^2 \ .
\label{eq:ComputingIandG2}
\end{split}
\end{equation}
From these equations, it follows that our perturbative
calculations can be restricted to second order, and we can
truncate the Hilbert space at the two-excitation manifold. To
compute the second-order correlation function $\gtau$, the
evolution operator
$
\mathcal{U}(t) = \mbox{exp}[{- \ii H_{\eff} t}]
$
has to be introduced:
\begin{equation}
\begin{split}
g^{(2)} (\tau)  = & \langle  \Ed ^- (0) \Ed^- (\tau) \Ed^+ (\tau) \Ed^+ (0) \rangle
/ I ^2
\\
= & \langle \psi | \Ed ^- \mathcal{U}^\dagger (\tau)  \Ed^-  \Ed^+ \mathcal{U}(\tau)  \Ed^+   | \psi \rangle
/ I ^2 \ ,
\end{split}
\end{equation}
where $ \Ed ^- \equiv  \Ed ^- (0)$, and the perturbative solution
of $| \psi \rangle$ up to second order is also used in the
calculation.

\section{Results and discussion} \label{sec:Results}

In the following, the theoretical framework presented in the
previous section is applied to study the coherence properties of
the light emitted by a collection of $N$ quantum emitters coupled
to either a nano- or a microcavity, with the distinction
introduced before. Our attention is focused on resonant coupling,
setting $\omega_{\C}, \omega_{\QE} \equiv \omega_0 = 3$ eV in all
cases (except in the section where the effects of spectral
detuning are explored). Beyond that, we have to
consider a specific set of parameters for each system. The
dissipation rate associated with the microcavity is taken to be
$\gamma_{\C} = 66 \ \mu$eV (16 GHz)~\cite{VuckovicNat2008}, which
corresponds to the spontaneous decay rate of a dipole moment
$\mu_{\C} = 3.1$ e$\cdot$nm \cite{NovotnyBook}. On the contrary,
substantial non-radiative losses are a distinctive feature of
plasmonic cavities \cite{MaierBook, SavastaACS2010}, hence we set
$\gamma_{\C} = 0.1$ eV (24 THz) for the nanocavity. This value
also incorporates the radiative losses corresponding to a dipole
with $\mu_{\C} = 19$ e$\cdot$nm, which mimics the cavity emission
\cite{GlezBallestero2015}. Regarding the emitters, we consider
those typically used in the experimental setups involving each
cavity. First, the quantum emitters usually located inside
dielectric cavities are characterized by negligible
non-radiative losses, thus $\gamma_{\QE}^{\nr} = 0$. In
particular, we choose semiconductor quantum dots with dipole
moment $\mu_{\QE} = 0.25$ e$\cdot$nm,  which corresponds to
a radiative decay rate $\gamma_{\QE}^{\rr} = 0.41 \ \mu$eV (0.10 GHz)
\cite{VuckovicNat2008}. Conversely, the quantum emitters
interacting with plasmonic nanocavities are considered to be
organic molecules with $\mu_{\QE} = 1$ e$\cdot$nm, and presenting
very low quantum yield. The specific rates chosen for these
emitters are $\gamma_{\QE}^{\nr} = 15$ meV
(3.6 THz) and $\gamma_{\QE}^{\rr} = 6 \ \mu$eV (1.5 GHz).

In our study, we assume that both open and closed cavities operate
at low temperature, which diminishes greatly the impact of
pure-dephasing processes in the dynamics of both quantum
dots~\cite{Thoma2016} and organic molecules~\cite{Rector1998}.
Accordingly, we have not included this decoherence mechanism in
our theoretical model. Moreover, from here on and for simplicity,
we consider that the external laser field $\boldsymbol{E}_{\rm L}$
is parallel to both the cavity and the quantum emitters dipole
moments. Note that this turns out to be the optimal configuration
to enter the strong coupling regime.

\subsection{Intensity and coherence} \label{subsec:R_IntensityCorrelation}

The features of the light emerging from these two configurations
are studied first as a function of the number of emitters. The
intensity and the zero-delay second-order correlation function for
the steady state are computed from \autoref{eq:ComputingIandG2}.
Although the numerical results displayed below depend on the
particular values of the parameters, the qualitative picture we
present is not bound to the specific configuration---on the
contrary, it remains the same when considering a wide range of
parameters describing realistic systems.

\subsubsection{Plasmonic nanocavities}

\begin{figure*}[htpb]
    \centering
    \includegraphics[width=\linewidth]{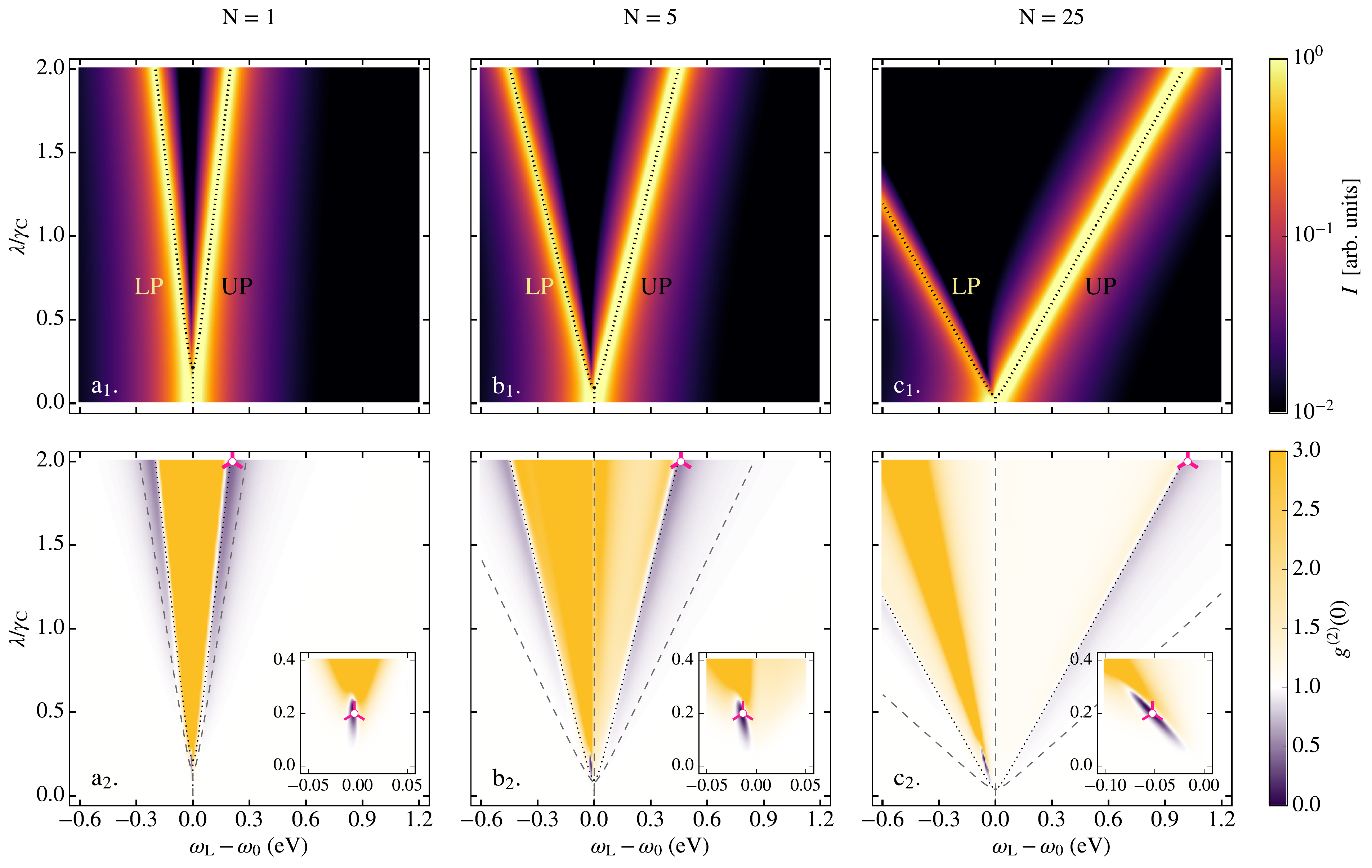}
    \caption{Scattering intensity $I$ (top row) and correlation function $g^{(2)}(0)$ (bottom row) versus laser detuning $\omega_{\LL} - \omega_0$ and coupling strength $\lambda$ (in units of the cavity decay rate $\gamma_{\C}$) for a system of $N = 1$ (a), 5 (b) and 25 (c) quantum emitters coupled to a plasmonic nanocavity. In these panels, dotted (dashed) lines plot the polariton frequencies (half-frequencies) in the one-excitation (two-excitation) manifold. Insets zoom into the low coupling region. Magenta marks indicate points whose $\gtau$ is plotted in \autoref{fig:tau} (a$_1$).}
    \label{fig:ms1_intg2}
\end{figure*}

We consider first the nanocavity, where the light reaching the
detector comes from both the quantum emitters and the cavity
itself. \autoref{fig:ms1_intg2} shows the scattering intensity $I$
(top row) and the zero-delay second-order correlation function
$g^{(2)}(0)$ (bottom row) for three different collections of
emitters: $N = 1$ (a), 5 (b), and 25 (c). Both magnitudes are
plotted as a function of the laser detuning $\omega_{\LL} -
\omega_0$ and the coupling strength $\lambda$, which is expressed
in units of the cavity decay rate $\gamma_{\C}$.

In all intensity maps two scattering maxima are observed, which
correspond to the polariton energies within the strong coupling
regime---that is, the eigenenergies of the dressed states in the
one-excitation manifold, the first-rung of the so-called {\it
    Tavis-Cummings ladder}. In this way, for each value of the
coupling strength we find two intensity peaks at laser frequencies
that match the lower (LP) and the upper (UP) polariton energies.
These dispersion curves are plotted in dotted lines overlapping
the maps, so that the correspondence is easily observed. Note that
these two maxima branches are also apparent within the weak
coupling regime, and thus this presence cannot be regarded as an
energy splitting. Its origin actually lies in a Fano-like
interference, appearing when two signals with very different
linewidths interact~\cite{Bryant2006, Bryant2008, SavastaPRL2010}.
The pronounced minimum in the scattering intensity is in this case
produced by the destructive interference between the cavity and
emitter emission.

Finally, observe the asymmetry between the two intensity maxima
branches: the one corresponding to the UP is distinctly brighter.
Note that the emission coming from each polariton can be described
from either the parallel (UP) or antiparallel (LP) superposition
of the dipole moments associated with the plasmon and the bright
mode of the emitter ensemble. Since the dipole moment
corresponding to the UP is larger, its emission is more intense.
This difference in the effective dipole moment between LP and UP
grows as $N$ increases, as a consequence of the greater collective
dipole moment of the ensemble. Hence the contrast between branches
becomes more pronounced for larger ensemble sizes.

We can get a better understanding from the analytical results
obtained thanks to the perturbative approach described above. The
expression for the scattered intensity reads:
\begin{equation}
I \propto \left|
\frac{  \tilde{\Delta}_{\C} \mu_{\QE}^2  + \dtQE \mu_{\C}^2/N - 2 \lambda   \mu_{\C} \mu_{\QE}}{\dtC \dtQE /N -  \lambda^2}
\right|^2 \ ,
\label{eq:anaOC_intensity}
\end{equation}
where the detunings of the laser frequency from both the cavity,
$\Delta_{\C}$, and the emitters, $\Delta_{\QE}$, are redefined  to
introduce the associated losses as $\dtC = \Delta_{\C} - \ii
\gamma_{\C}/2$ and $\dtQE = \Delta_{\QE}-\ii(\gamma_{\QE}^{\nr} +
N \gamma_{\QE}^{\rr})/2$. This expression confirms the origin of
the intensity maxima: the condition for which the denominator
vanishes, $\lambda^2 = \tilde \Delta_{\C} \tilde \Delta_{\QE}/N $,
gives us the dispersion of the LP and UP. Notice
the $\sqrt{N}$ dependence, characteristic scaling  of collective
coupling. In addition, this expression sheds light into the
asymmetry in the branches: the intensity behaves as $I \propto (1
\mp \sqrt{N} \mu_{\QE}/\mu_{\C})^2$ for the LP (upper sign) and UP
(lower sign) when the losses from both cavity and emitters are
neglected.

The bottom row of \autoref{fig:ms1_intg2} clearly shows areas of
super-Poissonian (yellow colored) as well as sub-Poissonian (blue
colored) statistics for all ensembles sizes. Focusing our
attention first in the single-emitter case ($N=1$), we distinguish
a main super-Poissonian area located between the LP and UP
energies (again depicted as dotted lines) for all coupling values.
Close to these polariton frequencies, but still far from the
two-excitation eigenenergies (depicted as dashed lines), regions
of sub-Poissonian light are found, being more pronounced as the
coupling strengthens. These correspond to the well-known {\it
    photon blockade} effect, where the presence of an excitation in
the system prevents the absorption of a second photon at certain
frequencies due to the anharmonicity of the energy ladder. Apart
from these three stripes, we find another area of sub-Poissonian
emission that is enlarged in the corresponding inset. It lies
around the resonant frequency $\omega_{\LL} = \omega_{0}$. The
mechanism behind it was addressed theoretically in the context of
dielectric microcavities~\cite{Savona2010}, the so-called {\it
    unconventional antibunching}. Here, we employ the term  {\it
    interference-induced correlations}, since it is the destructive
interference among possible decay paths that produces the
suppression of two-photon processes and hence the drop of $\gz$
below one~\cite{Ciuti2011, Ciuti2013}. In the following section,
these two different types of sub-Poissonian light are discussed in
further detail.

The statistical features observed for single emitters are also
present for larger $N$. As we already pointed out in
Ref.~\cite{Rocio2017}, photon correlations arising at the
single-emitter level remain, and can even be enhanced as the
ensemble size increases. This is observed in the panels
corresponding to $N=5$ and $N=25$ in \autoref{fig:ms1_intg2}. The
area of bunched light remains between the one-excitation
eigenenergies, although it tends to approach the LP branch when
the number of emitters increases. This tilt is also observed for
the interference-induced correlation area: the region of negative
correlations shifts towards the LP energy, while values for the
function $\gz$ below one are still achieved within the same
coupling range as for the single-emitter case. Note that by
increasing further the number of emitters, the system eventually bosonizes
(that is, it yields $\gz=1$). The quantum character of
the emitted light is then lost. Focusing now on the region
associated with the photon blockade effect, we observe that the
minimum following the UP branch is deeper than the LP one.
Nevertheless, it is apparent how both fade for larger sizes of the
emitter ensemble. Indeed, for $N=25$ there are a wide range of
coupling strengths where this effect is not observable. On the
contrary, for moderate values of the coupling the dip
corresponding to interference effects is not only present, but
becomes quite pronounced.

From our perturbative approach we can also obtain the analytical
expression for the correlation function $g^{(2)}(0)$ corresponding
to an open nanocavity:
\begin{widetext}
\begin{equation}
\begin{split}
g^{(2)}(0) = & \left| 1 - \frac{1}{N}
\left( \frac{\dtC \mu_{\QE} - \lambda \mu_{\C}}
{ \dtC \mu_{\QE}^2  + \dtQE \mu_{\C}^2/N  - 2 \lambda  \mu_{\C} \mu_{\QE}}\right)^2
\right. \\
& \left.
\frac{( \dtQE + \ii N \gamma_{\QE}^{\rr}/2)[\dtC \dtQE \mu_{\QE}^2 +(\dtC \mu_{\QE} - \lambda \mu_{\C} )^2 -  N \lambda^2 \mu_{\QE}^2]}
{( \dtQE+\ii \gamma_{\QE}^{\rr}/2)(\dtC^2  +  \dtC  \dtQE -N \lambda^2) - \dtC (N-1)\lambda^2}
\right|^2 \ ,
\end{split}
\label{eq:anaOC_g2}
\end{equation}
\end{widetext}
where, again, we have made use of the redefined detunings $\dtC$
and $\dtQE$. We observe that, as expected, $\gz \rightarrow 1$
when $N$ tends to infinity for a fixed value of the coupling
strength---so the expression does recover the bosonization limit.
Notice also that the denominator of the first term coincides with
the numerator of the analytical expression for the intensity
(\autoref{eq:anaOC_intensity}), and the vanishing condition for
the denominator of the second term yields the polaritons of the
second-rung of the Tavis-Cummings ladder.

\begin{figure*}[ht]
    \centering
    \includegraphics[width=15cm]{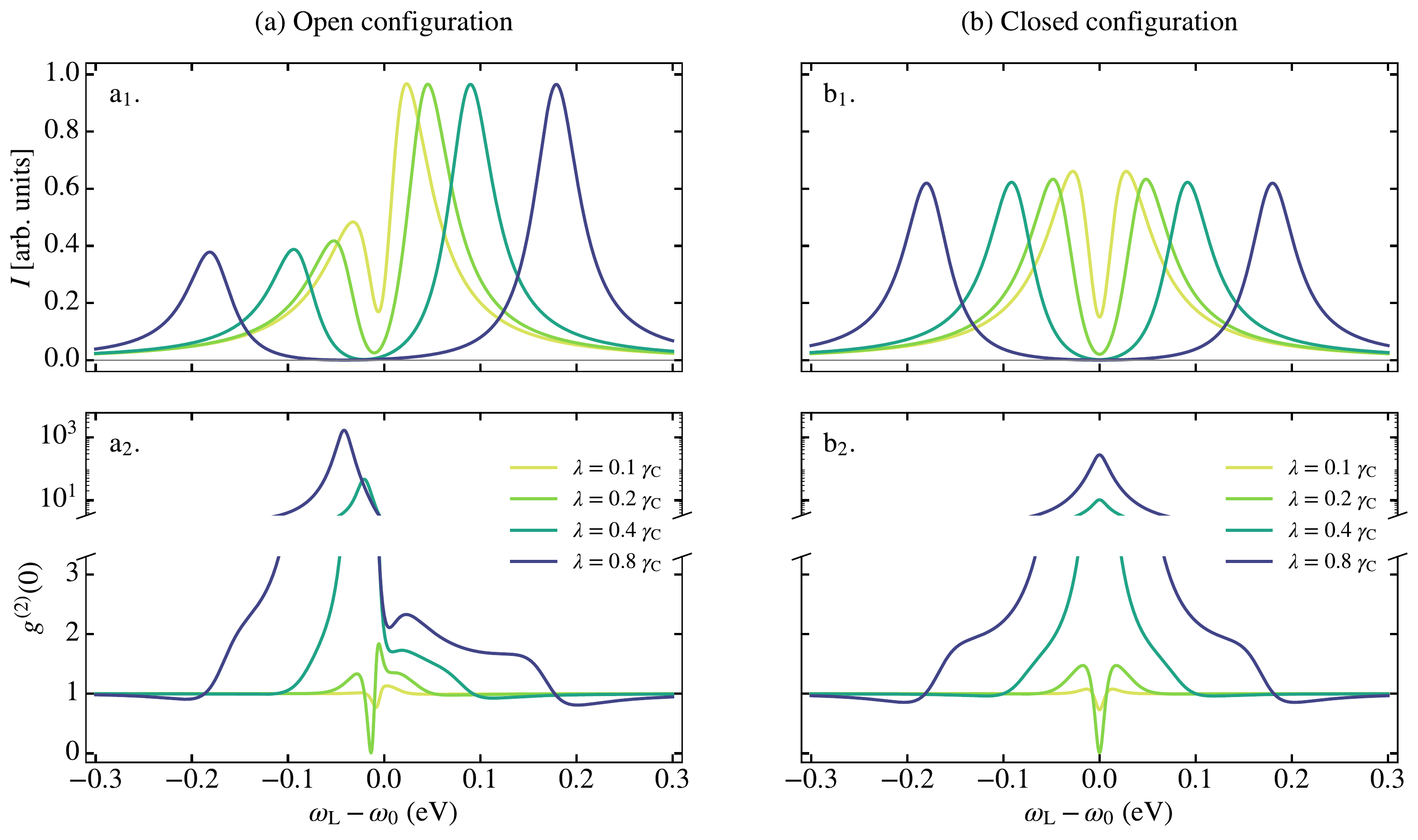}
    \caption{Scattering intensity $I$ (top row) and correlation function $g^{(2)}(0)$ (bottom row) versus laser detuning $\omega_{\LL} - \omega_0$ for a system of $N = 5$ quantum emitters coupled to a plasmonic nanocavity for various coupling strengths $\lambda$ with (a) and without (b) considering the pumping to and the emission from the quantum emitters.}
    \label{fig:ms1_comp}
\end{figure*}

Considering the same cavity and emitters as in the plasmonic case,
we can explore the changes introduced when quantum emitters are
not directly pumped by the laser and only the radiation coming
from the cavity is registered at the detector. This mimics the
setup of a closed microcavity for the parameter values distinctive
of a plasmonic nanocavity. Results for the intensity $I$ and the
correlation function $g^{(2)} (0)$ for different values of the
coupling strength $\lambda$ are shown in \autoref{fig:ms1_comp}
for these two situations: with (a) and without (b) considering
both the pumping and the emission associated with the quantum
emitters. To make the comparison, we present the particular case
of a collection of $N=5$ quantum emitters. Therefore, the lines
appearing in the left-hand column of \autoref{fig:ms1_comp} are
just cuts of the maps (b$_1$) and (b$_2$) of
\autoref{fig:ms1_intg2} at four particular values of the coupling
strength.

First, we notice that the asymmetry in the two intensity peaks is
removed when considering only the emission from the cavity. As
commented before, for an open nanocavity the effective dipole
moment of the LP and the UP are respectively the parallel and
antiparallel superpositions of those of the cavity and the quantum
emitters. This makes the emission of the UP brighter, and it also
introduces a noticeable dependence on the number of emitters. On
the contrary, in this new configuration both polaritons radiate
with the same associated dipole moment (corresponding to the
cavity, which is the same regardless the ensemble size), and the
associated emission is thus identical. Notice also that the
positions of the intensity peaks are the same for open and closed
configurations, because the polariton energies do not change.

The symmetry observed in the intensity patterns is also kept in
the correlation function $g^{(2)}(0)$. Apart from this difference,
the main features in the statistics are kept from the open case,
namely: first, around the zero value of the laser detuning
($\omega_{\LL} = \omega_0$) there exists a dip for reduced
coupling strength whereas a maximum is developed as the
interaction increases; and second, for frequencies near the
one-excitation polaritons, the photon blockade effect is
observable. From these cuts in \autoref{fig:ms1_comp}(a$_2$), we
confirm that the minimum following the UP branch is the deepest.

\subsubsection{Dielectric microcavities}

\begin{figure*}[ht]
    \centering
    \includegraphics[width=\linewidth]{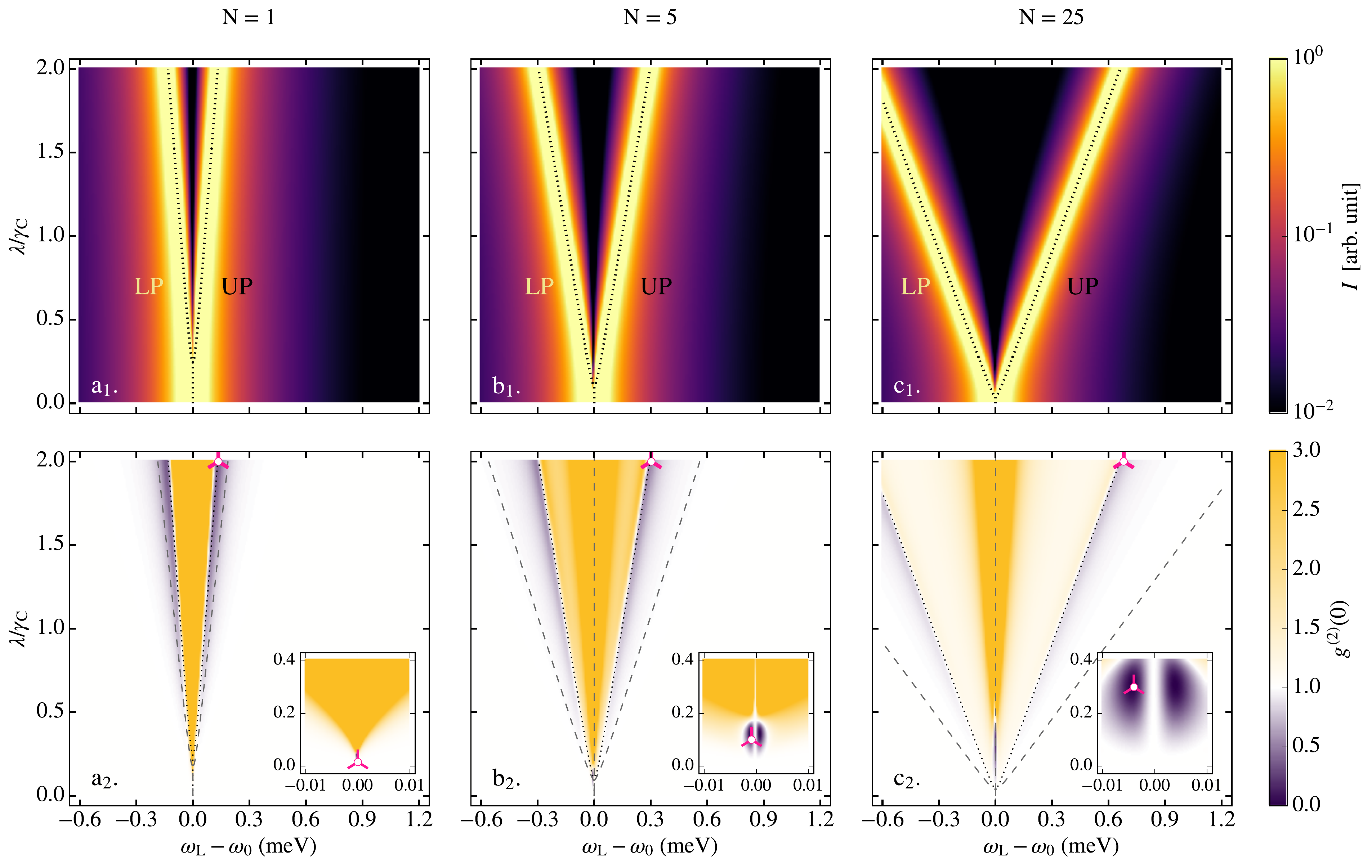}
    \caption{Scattering intensity $I$ (top row) and correlation function $g^{(2)}(0)$ (bottom row) versus laser detuning $\omega_{\LL} - \omega_0$ and coupling strength $\lambda$ (in units of the cavity decay rate $\gamma_{\C}$) for a system of $N = 1$ (a), 5 (b) and 25 (c) quantum emitters coupled to a dielectric microcavity. In these panels, dotted (dashed) lines plot the polariton frequencies (half-frequencies) in the one-excitation (two-excitation) manifold. Insets zoom into the low coupling region. Magenta marks indicate points whose $\gtau$ is plotted in \autoref{fig:tau} (b$_1$).}
    \label{fig:ms2_intg2}
\end{figure*}

Now we consider the typical configuration of dielectric
microcavities, where only the cavity mode is pumped and direct
emission from the quantum emitters does not take place (closed
configuration). A study similar to the previous section is carried
out, determing the intensity $I$ and the zero-delay second-order
correlation function $g^{(2)} (0)$. The results are shown in
\autoref{fig:ms2_intg2} for the same three cases: $N = 1$ (a), 5
(b), and 25 (c) quantum emitters placed inside the cavity.

The intensity panels reveal again the presence of the two
polaritons when entering the strong coupling regime. The energies
corresponding to the dressed states in the one-excitation manifold
are plotted in dotted lines, and they overlap the intensity peaks.
Nevertheless, in contrast to the nanocavity configuration, these
two intensity maxima are symmetric and barely vary their height as
the number of emitters increases. As we have commented in the
previous section, this is due to the fact that only the emission
from the cavity is registered, so the effective radiating dipole
moment is always the same.

This underlying symmetry is also revealed in the analytical
expression of the intensity, computed from the perturbative
approach:
\begin{equation}
I \propto \left|
\frac{\dtQE \mu_{\C}^2 /N}{\dtC \dtQE /N -  \lambda^2}
\right|^2 \ .
\end{equation}
This expression can be reproduced from that corresponding to nanocavities, \autoref{eq:anaOC_intensity}, just by considering the limit $\mu_{\QE} \rightarrow 0$. We observe that the denominator remains unchanged, so its vanishing condition give us again the energy dispersion for the polaritons and hence the position of the intensity maxima.

The results for the correlation function $\gz$ depicted in the
bottom row of \autoref{fig:ms2_intg2} show a similar pattern to
the ones found for nanocavities (\autoref{fig:ms1_intg2}),
although incorporating the symmetry already expected. Around the
zero laser detuning ($\omega_{\LL} = \omega_0$) and for
intermediate (or large) coupling strengths we find
super-Poissonian statistics (yellow colored). These panels also
show the two types of sub-Poissonian emission (blue colored)
previously observed: the associated with the phenomenon of photon
blockade, as well as that related to destructive interference. The
position of the eigenenergies corresponding both to the one- and
two-excitation manifolds are plotted in dotted and dashed lines,
respectively, overlapping the correlation maps. The frequencies
where the photon blockade effect occurs are easily relatable next
to the dotted lines. Nevertheless, it is in the other area of
sub-Poissonian emission---the one associated with interference
effects---where the main difference between open and closed
cavities appears. Apart from the spectral symmetry already
discussed, the development of two dips instead of a single one is
the most apparent feature. For lower values of the coupling
strength, we find $\gz=1$ at zero detuning, while on both sides of
this frequency, a window with sub-Poissonian emission is visible.
The presence of this double dip pattern disappears when
introducing non-radiative losses associated with the quantum
emitters.

The evolution of correlations as the number of emitters increases
differs from the open nanocavity case. Apart from the fact that
symmetry modifies the laser frequencies at which the different
regions are achieved (for a specific value of the coupling
strength), the main variation concerns the sub-Poissonian emission
caused by destructive interference. These areas are enlarged in
the insets of \autoref{fig:ms2_intg2}. As the ensemble size
increases, the parameter ranges in which we find $\gz <1$ clearly
widen. Therefore, it is possible to obtain antibunched emission
for a specific coupling strength just by increasing the number of
emitters. Beyond a particular $N$, the system tends to reach the
bosonization limit, where $\gz = 1$. The onset of this regime
depends on the coupling strength between cavity and emitters and,
as seen in \autoref{fig:ms2_intg2}(c$_2$), for $N=25$ emitters we
still find significant negative correlations for a wide interval
of coupling values. Note that the photon blockade region does not
endure so long and it practically disappears for a few emitters
within this coupling range.

There exists a major aspect, not mentioned before, that should be
highlighted: the range of laser detunings at which this
non-classical behaviour is found is of the order of meV. Notice
that the energy scale in \autoref{fig:ms2_intg2} differs in three
orders of magnitude from the one corresponding to nanocavities,
\autoref{fig:ms1_intg2}. Therefore, the spectral robustness and
accessibility of the antibunched regions is significantly
different in open and closed cavities, specially in the case of
interference-induced negative correlations. We anticipate that the
spectrally broad (narrow) nature of photon correlations in
plasmonic nanocavities (dielectric microcavities) implies a
faster (slower) temporal evolution of $\gtau$.

Finally, to gain insight into the coherence properties discussed
above, we present the analytical expression for the correlation
function $g^{(2)}(0)$:
\begin{widetext}
\begin{equation}
g^{(2)}(0) =  \left| 1 - \frac{1}{N}
\left( \frac{\lambda}
{\dtQE / N}\right) ^2
\frac{ ( \dtQE + \ii N \gamma_{\QE}^{\rr}/2) \lambda^2}
{( \dtQE+\ii \gamma_{\QE}^{\rr}/2)(\dtC^2  +  \dtC  \dtQE -N \lambda^2) -\lambda^2 \dtC (N-1)}
\right|^2 \ ,
\end{equation}
\end{widetext}
which can be also obtained by taking $\mu_{\QE} \rightarrow 0$ in
\autoref{eq:anaOC_g2}. Again, this expression yields $\gz=1$ when
$N \rightarrow \infty$, so the classical behaviour is recovered in
this limit. Note as well that we find $\gz = 1$ at the resonant
frequency $\omega_{\LL} = \omega_0$ when all losses are neglected.

\subsection{Two different mechanisms leading to sub-Poissonian light} \label{subsec:R_TwoMechanisms}

\begin{figure*}[ht]
    \centering
    \includegraphics[width=\linewidth]{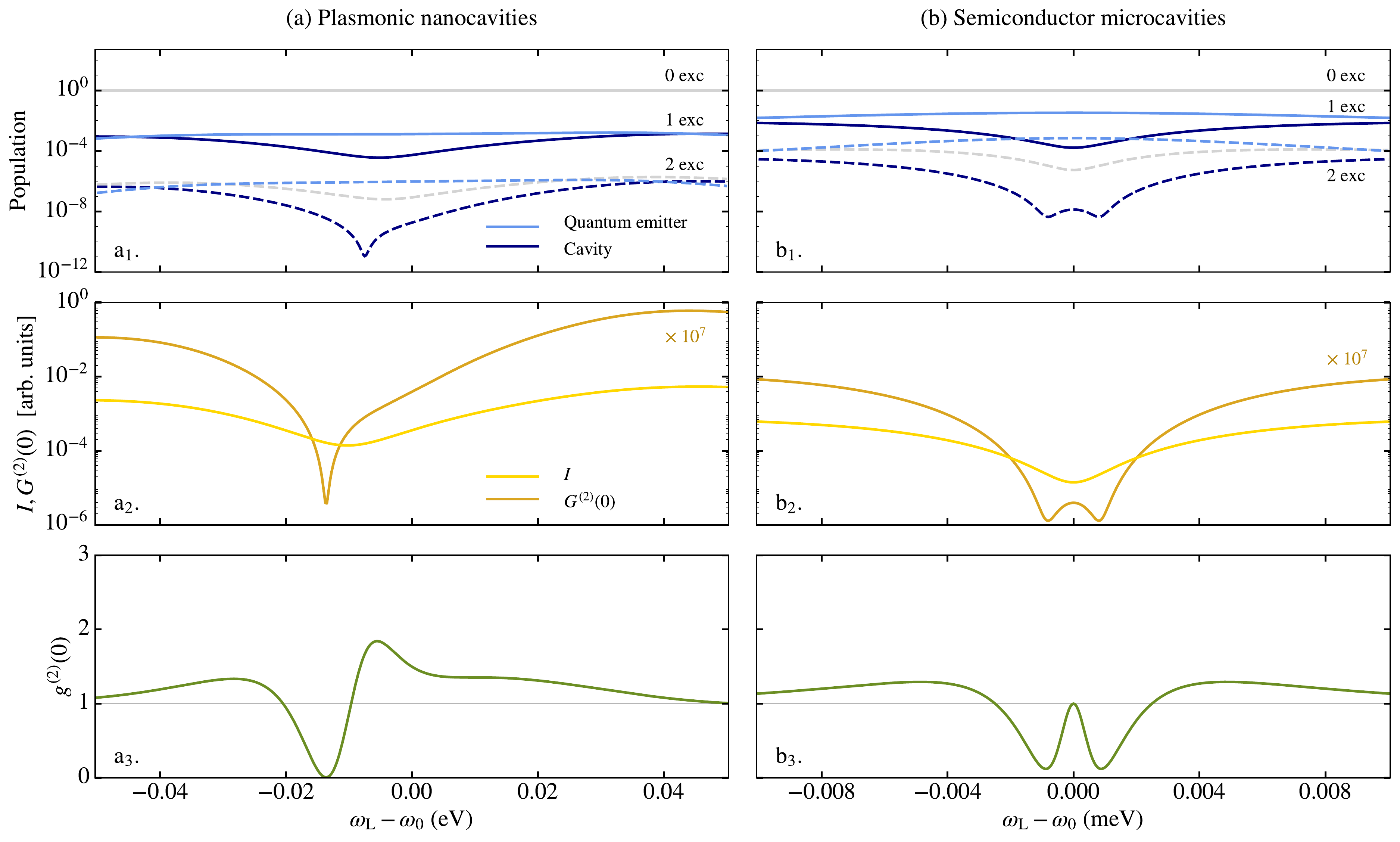}
    \caption{Population (in the cavity-emitters basis), intensity $I$, and correlation functions $G^{(2)}(0)$ and  $g^{(2)}(0)$ versus laser detuning $\omega_{\LL} - \omega_0$ for a system of $N=5$ quantum emitters coupled to a nano- (a) and a micro- (b) cavity at coupling strengths $\lambda / \gamma_{\C} = 0.2$ and $0.1$ respectively (for which the interference-induced correlations appear).}
    \label{fig:pop_interference}
\end{figure*}

Studying photon correlations in coupled systems, we have
identified two types of sub-Poissonian emission appearing in both
nano- and microcavities. In order to shed light into their
different nature, we proceed in this section to examine their
emergence in more detail.  In \autoref{fig:pop_blockade}, the
focus is on the photon blockade effect---which takes place close
to the polarition energies of the one-excitation manifold---,
whereas in \autoref{fig:pop_interference} we study the
sub-Poissonian emission associated with destructive
interference---which appears for moderate coupling strength in
the region of zero-detuning. The population, the intensity $I$, and the
correlation functions $G^{(2)}(0)$ and $g^{(2)}(0)$ are plotted as
a function of the laser detuning $\omega_{\LL} - \omega_0$ for
nano- (left-hand side panels) and micro- (right-hand side panels)
cavities at specific coupling strengths to explore these
processes.

\subsubsection{Photon blockade}

\begin{figure*}[ht]
    \centering
    \includegraphics[width=\linewidth]{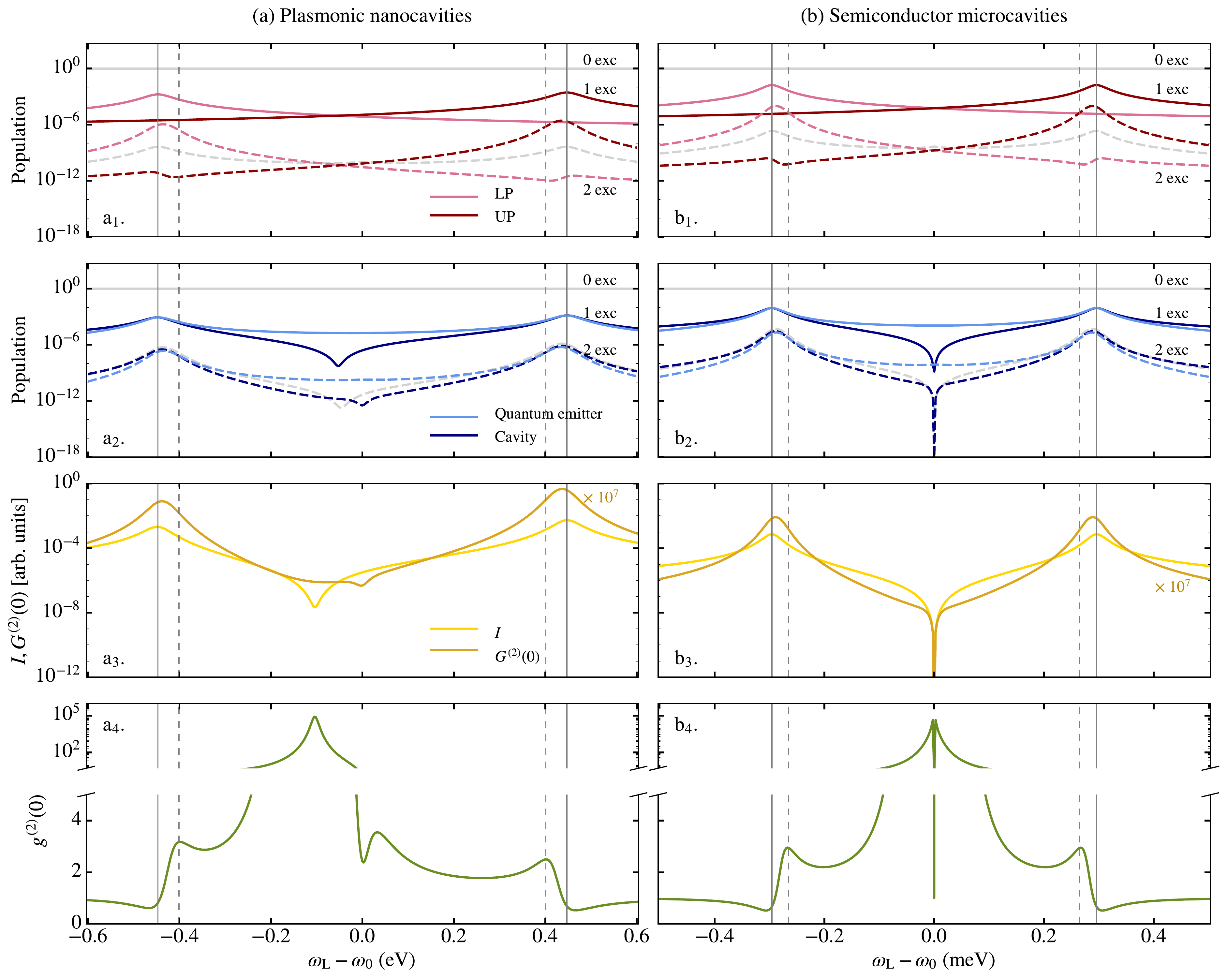}
    \caption{Population---in the polariton basis (first row) and in the cavity-emitters basis (second row)---, intensity $I$, and correlation functions $G^{(2)}(0)$ and  $g^{(2)}(0)$ versus laser detuning $\omega_{\LL} - \omega_0$ for a system of $N=5$ quantum emitters coupled to a nano- (a) and a micro- (b) cavity at coupling strength $\lambda / \gamma_{\C} = 2$ (for which the photon blockade effect appears). In these panels, continuous vertical grey lines indicate the polariton frequencies in the one-excitation manifold, while the dashed ones represent the energy differences between the state of one LP and one UP (belonging to the two-excitation manifold) and the state with either one LP (left dashed line) or one UP (right dashed line).}
    \label{fig:pop_blockade}
\end{figure*}

The sub-Poissonian emission due to the {\it photon blockade
    effect} originates from the anharmonicity of the Tavis-Cummings
ladder, as we have commented before. When the laser has an energy
close to that of one of the polaritons at the one-excitation
manifold, the population of this particular hybrid state
increases. In panels (a$_1$) and (b$_1$) of
\autoref{fig:pop_blockade}, populations are plotted in the basis
of the dressed states. There, the continuous coloured lines,
corresponding to the population of the UP (dark pink) and the LP
(light pink), experience an increase when the laser frequency is
tuned to be in the vicinity of the corresponding polariton
frequency (continuous vertical grey lines). Nevertheless, for
these specific energies, the laser is out of resonance for
promoting the state from the one- to the two-excitation manifold
(dashed vertical grey lines depict these energy differences). This
diminishes the probability of emission of two simultaneous photons,
leading to sub-Poissonian statistics.
These panels also show, in dashed coloured lines, the populations
of the states belonging to the two-excitation manifold: two LPs
(light pink), one LP and one UP (very light grey) and two UPs
(dark pink). Note that the maxima of these curves are not located
exactly at the polariton frequencies. They are slightly shifted as
a consequence of the energy differences between the one- and the
two-excitation manifolds.

When the populations are expressed in terms of the cavity and
emitter states, panels (a$_2$) and (b$_2$) of
\autoref{fig:pop_blockade}, all curves belonging to the same
subspace (continuous or dashed lines for the one- and
two-excitation manifolds respectively) seem to converge to the
same value at the frequencies where the photon blockade phenomenon
takes place (that is, near the polariton frequencies). Apart from
that, we observe that there exist two clear minima in the
population curves corresponding to the state with one (continuous
dark blue line) and two (dashed dark blue line) excitations in the
cavity. Each of them has a replica in one of the curves depicted
in panels (a$_3$) and (b$_3$) of \autoref{fig:pop_blockade}. This
is especially visible for the nanocavities where these two minima
do not coincide. Indeed, the intensity (yellow line) and the $\Gz$
(ochre line) functions reproduce the form of the populations of
the states with one and two excitations in the cavity mode
respectively. The origin of this correspondence is clear for the
closed configuration (as only the emission from the cavity is
detected). For the open one, it results from the fact that the
dipole moment of the cavity is greater than the collective dipole
of the emitter ensemble. Thus, the former contributes the most to
the emitted light (for a reduced number of emitters). Note that
intensity accounts for one-photon processes, while $\Gz$ reflects
from two-photon processes instead (\autoref{eq:ComputingIandG2}).

The intensity plots reflect the presence of the polariton energies
as well---each scattering peak coincides with a maximum in the
polariton population and, naturally, with the position of the
polariton energy. The intensity minima are certainly located
between the two polariton energies, far from resonance. The fact
that the maxima in $G^{(2)}(0)$ are shifted from those in the
scattered intensity provokes the characteristic shape in the
normalized second-order correlation function, shown in panels
(a$_4$) and (b$_4$) of \autoref{fig:pop_blockade}. Values of the
$\gz$ function below one are located close to the polariton
energies (vertical continuous grey lines), whereas there appear
two relative maxima at laser frequencies that match energy
differences between the one- and the two-excitation manifold
(vertical dashed grey lines). For the nanocavity, the different
positions of $I$ and $G^{(2)}(0)$ minima in (a$_3$) leads to
maxima and a minima in $\gz$ near resonance, although the emission
is always super-Poissonian in this frequency window.

\subsubsection{Interference-induced correlations}

The decrease of $\gz$ below one is referred to as
interference-induced correlations when its origin cannot be
explained in terms of the energy levels as done for the photon
blockade effect. On the contrary, it is produced by the
destructive interference between different available decay
paths~\cite{Savona2010, Ciuti2011, Ciuti2013}.

The population curves, panels (a$_1$) and (b$_1$) in
\autoref{fig:pop_interference}, reveal that it is a decrease in
the population of the state corresponding to two cavity-mode
excitations (dashed dark blue lines) that produces the mimimum in
the $G^{(2)} (0)$ function. It is then transferred to the
normalized $\gz$ and, consequently, there appears sub-Poissonian
statistics in the vicinity of this laser frequency.  This
correspondence between cavity population and correlations is
observed in both types of cavities, although there exists a
difference between them: whereas only one dip takes place in
nanocavities, two of them emerge in the case of microcavities.
Notice that this behaviour differs from the photon blockade
mechanism, where a related fall in the population of the state
with two cavity-mode excitations is not observed (on the contrary,
as previously pointed out, all populations seem to converge to the
same value).

The intensity and correlation function $\Gz$ are depicted in
panels (a$_2$) and (b$_2$) in \autoref{fig:pop_interference}. As
in the previous case, these curves clearly follow the shape of the
populations associated with the states corresponding to one
(continuous dark blue lines) and two (dashed dark blue lines)
excitations in the cavity mode, respectively. The different
position of the minima for these two magnitudes is again
responsible for the shape of the $\gz$ function. Nevertheless, now
values below one are reached (the interference-induced photon
correlations). In \autoref{fig:pop_blockade}, the minimum in the
$G^{(2)}$ did not lead to sub-Poissonian statistics, although it
did correspond to a minimum in $\gz$. Note that this fall was not
so abrupt when compared with the intensity dip.

\begin{figure*}[ht]
    \centering
    \includegraphics[width=\linewidth]{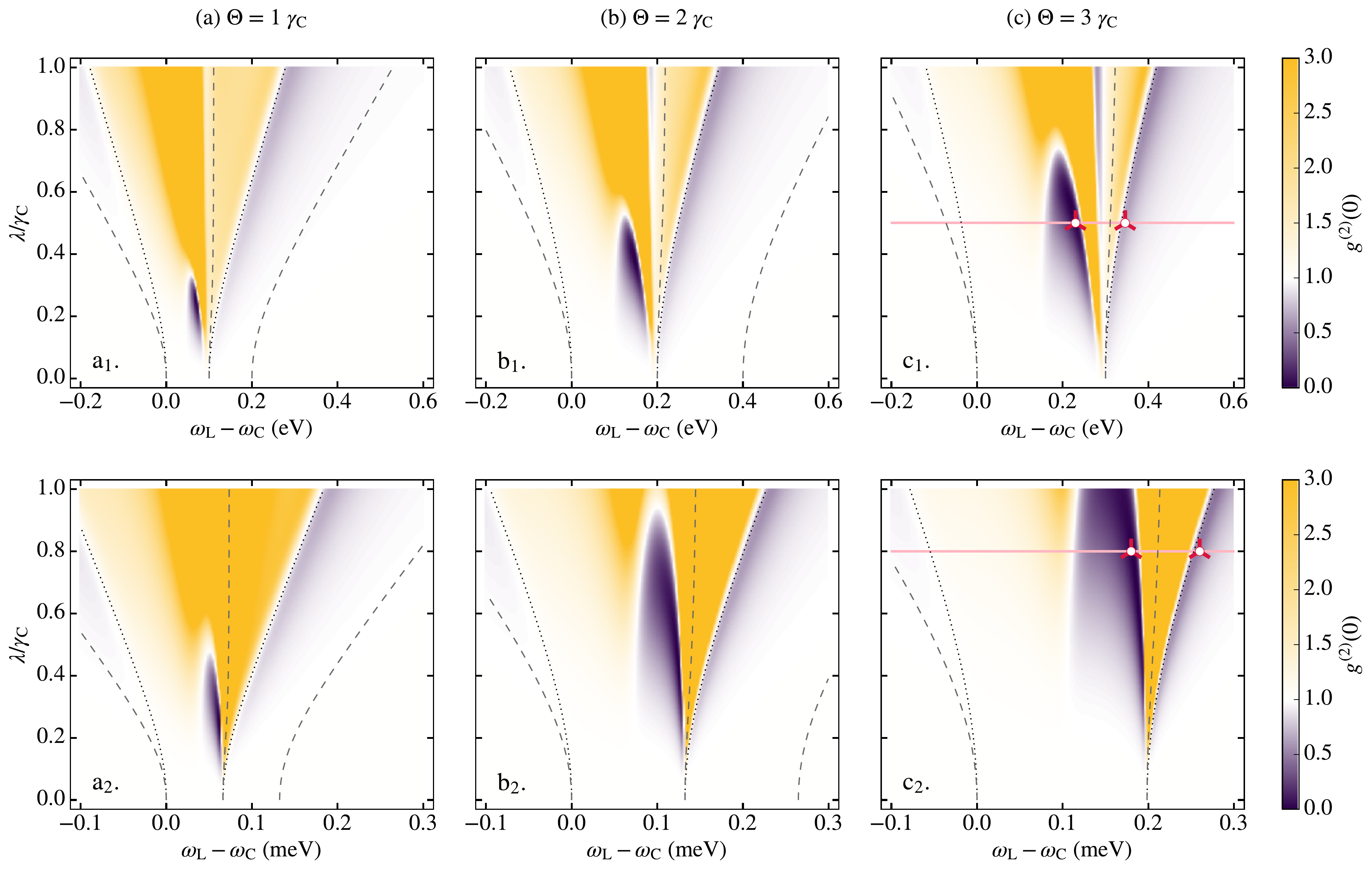}
    \caption{Correlation function $g^{(2)}(0)$ versus laser detuning $\omega_{\LL} - \omega_{\C}$ and  coupling strength $\lambda$ (in units of the cavity decay rate $\gamma_{\C}$) for a system of $N=5$ quantum emitters coupled to a nano- (top row) and a micro- (bottom row) cavity for various values of the detuning between cavity and emitters, $\Theta$ = 1 $\gamma_{\C}$ (a), 2 $\gamma_{\C}$ (b), and 3 $\gamma_{\C}$ (c). In these panels, dotted (dashed) lines plot the polariton frequencies (half-frequencies) in the one-excitation (two-excitation) manifold. Horizontal pink lines and magenta marks in (c) panels indicate points whose $\gz$ is plotted in \autoref{fig:tauDet}.}
    \label{fig:det}
\end{figure*}

\subsection{Effect of detuning between cavity and emitters frequencies on the correlation function $\boldsymbol{\gz}$} \label{subsec:R_Detuning}

By means of the introduction of detuning between cavity and
emitters frequencies, the parameter range in which sub-Poissonian statistics
emerges can be enlarged---the spectral window becomes wider, and
stronger couplings are required~\cite{Radulaski2017}. This is the
tendency we observe in \autoref{fig:det}, where the correlation
function $\gz$ is plotted versus the laser detuning $\omega_{\LL}
- \omega_{\C}$ and the coupling strength $\lambda$ for various
values of the detuning $\Theta \equiv \omega_{\QE} - \omega_{\C}$.
There, the emitter frequencies $\omega_{\QE}$ vary while the
cavity mode resonance is always fixed to be $\omega_{\C} =
\omega_{0} \equiv  3$ eV. The case considered is that composed of $N=5$
quantum emitters coupled to either a nano- (top row) or a micro-
(bottom row) cavity, hence these would correspond to panels
(b$_2$) from \autoref{fig:ms1_intg2} and \autoref{fig:ms2_intg2},
respectively, if no detuning were present (that is, $\Theta = 0$).

\autoref{fig:det} shows that, effectively, for both types of
cavities the region with interference-induced correlations spreads
as the difference in energy between cavity and emitters increases,
although this effect is more pronounced in microcavities. Via
detuning, the range of laser frequencies for which sub-Poissonian
emission is attainable broadens---it extends over a frequency
window with a width of almost half the detuning. Focusing now on
the vertical axis, we observe that for a particular value of the
coupling strength, it is possible to have $\gz < 1$ near the
resonant frequency just by increasing the detuning between cavity
and emitters. Furthermore, note that the introduction of detuning
makes it possible to achieve lower values of $\gz$, whereby
improving the quantum character of the emitted light. This is also
true for the photon blockade effect following the UP---since this
is the dressed state with a greater emitter contribution in this
case---, which deepens. For a better visualization, the energies
corresponding to the eigenvalues of the dressed states are plotted
in dotted (one-excitation manifold) and dashed (two-excitation
manifold) lines in all panels. For both nano- and microcavities,
the photon blockade effect reinforces near the UP, whereas it
fades at the LP. For instance, when $\Theta = 3 \gamma_{\C}$ the
photon blockade effect following the lower branch disappears---no
sub-Poissonian emission takes place in its surroundings.
Note that varying the sign of the detuning, the roles of UP and LP
are exchanged.

\begin{figure*}[ht]
    \centering
    \includegraphics[width=15cm]{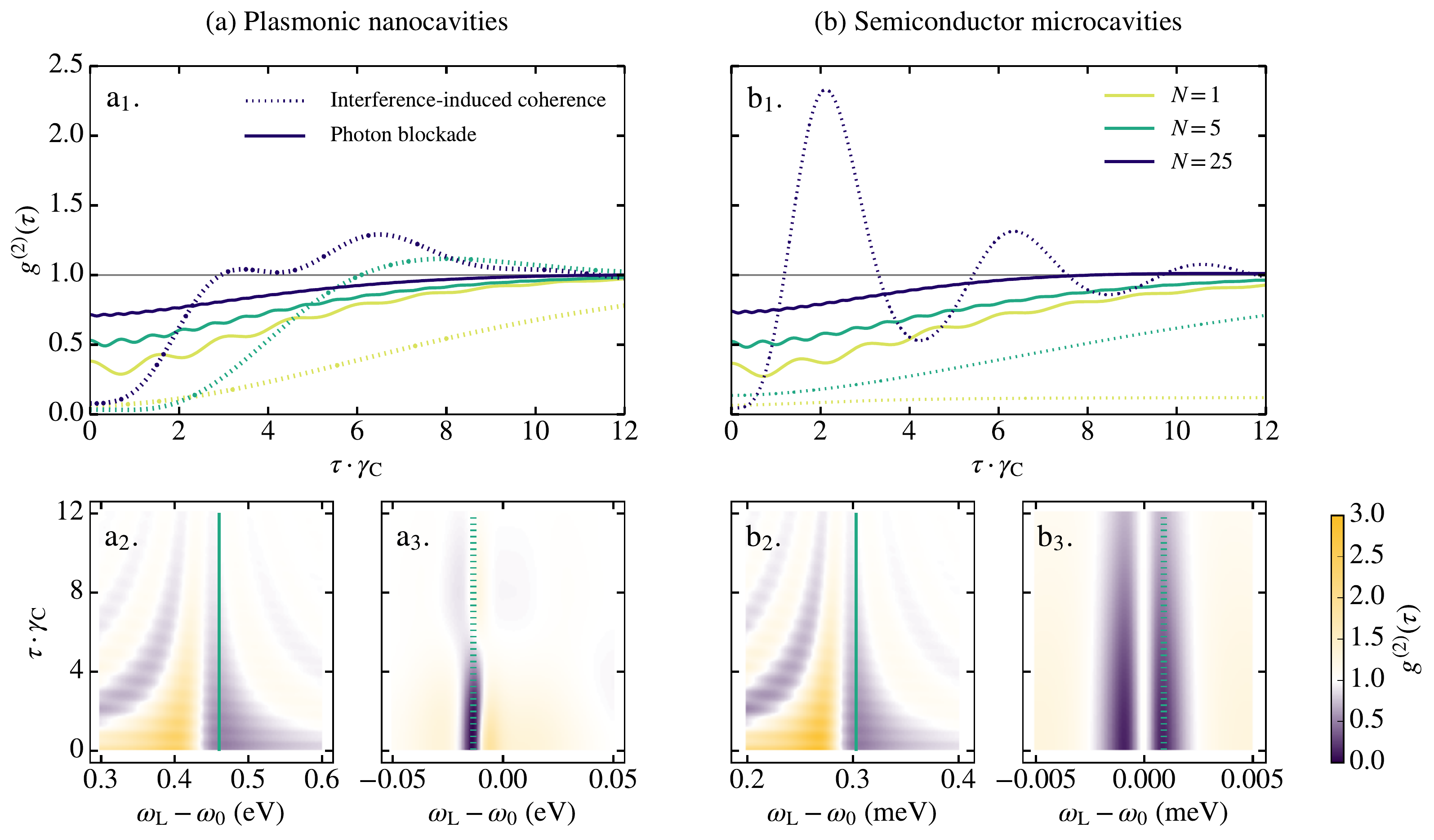}
    \caption{Correlation function $\gtau$ for $N$ quantum emitters interacting with either a nano- (a) or a micro- (b) cavity at resonance ($\Theta=0$). In top panels, $\gtau$ is plotted versus the time delay $\tau$ (in units of the cavity time $1/\gamma_{\C}$) for three different ensemble sizes ($N =1$, 5 and 25). Two different configurations (which correspond to the points indicated in \autoref{fig:ms1_intg2} and \autoref{fig:ms2_intg2}) are selected. Continuous lines are used for the antibunching related to the photon blockade effect while dotted lines are used for interference-induced correlations. In bottom panels, $\gtau$ is plotted versus laser detuning $\omega_{\LL} - \omega_0$ and time delay $\tau$ (also in units of $1/\gamma_{\C}$) for an ensemble of $N=5$ emitters and selecting two different coupling strengths in each case: $\lambda = 2 \gamma_{\C}$ (a$_2$) and $0.2 \gamma_{\C}$ (a$_3$) for plasmonic nanocavities and $\lambda = 2 \gamma_{\C}$ (b$_2$) and $0.1 \gamma_{\C}$ (b$_3$) for dielectric microcavities. In these bottom panels, vertical green lines correspond to the curves for $N=5$ depicted at the top. }
    \label{fig:tau}
\end{figure*}

Regarding the region with interference-induced correlations, there
exists a particularity for the microcavity that is worth mentioning:
now we only observe one prevailing dip, instead of two (as it was
for the zero detuning case, $\Theta = 0$). As a consequence of the
loss of symmetry, the dip closer to the emitter frequency becomes
narrower, and the other one widens, when the detuning increases.
This makes the patterns observed for $\gz$ at a specific detuning
$\Theta$ quite similar for nano- and microcavities. Nevertheless,
the energy range is very different---note that the coupling
strength $\lambda$ is given in units of
the decay rate $\gamma_{\C}$, and the values corresponding to plasmonic
($\gamma_{\C} \sim 0.1$ eV) and dielectric ($\gamma_{\C} \sim
0.1$ meV) cavities differ by around three orders of magnitude (as do
the laser detunings).

\subsection{Dependence of the correlation function $\boldsymbol{\gtau}$ on the time delay $\tau$} \label{subsec:R_Tau}

In this section, we study the behaviour of the second-order
correlation function $\gtau$ at non-zero time delays $\tau$ for
various configurations displaying sub-Poissonian statistics ($\gz < 1$).
This allows to resolve whether the emitted light is actually
antibunched ($\gz < \gtau$). In top panels of \autoref{fig:tau},
we plot $\gtau$ as a function of $\tau$ (in units of the cavity
lifetime $1/\gamma_{\C}$) for an ensemble of $N$ quantum emitters
interacting with either a nano- (a) or a microcavity (b) when
there is no detuning between them ($\Theta = 0$). We consider
three ensemble sizes $N=1$ (yellow lines), 5 (green lines) and 25
(blue lines), and select two different configurations for each
case: one belonging to the photon blockade area (continuous lines)
and the another displaying sub-Poissonian statistics due to
quantum interference effects (dotted lines). All configurations
are indicated in \autoref{fig:ms1_intg2} and
\autoref{fig:ms2_intg2} through magenta marks.

Focusing first on the continuous lines (photon blockade), we
observe that the correlation function approaches one almost
monotonically as the time interval $\tau$ increases, hence we can
talk properly of photon antibunching. Nevertheless, there exist
some oscillations whose amplitude diminishes as the ensemble size
increases. Remarkably, there is practically no difference between
the behaviour for nano- and microcavities once the time scale is
normalized by the cavity decay time $1/\gamma_{\C}$---in both
cases, the time evolution follows the same tendency, and the
degree of correlation reached from both effects is similar. Note
that all configurations have been chosen for a coupling strength
$\lambda/\gamma_{\C} = 2$ for both types of cavities, although the
laser detuning varies in order to consider the minimum of the
$\gz$ attainable at this coupling. These two sets of curves also
highlight that the degree of coherence is quickly lost when
increasing $N$.

\begin{figure*}[ht]
    \centering
    \includegraphics[width=13cm]{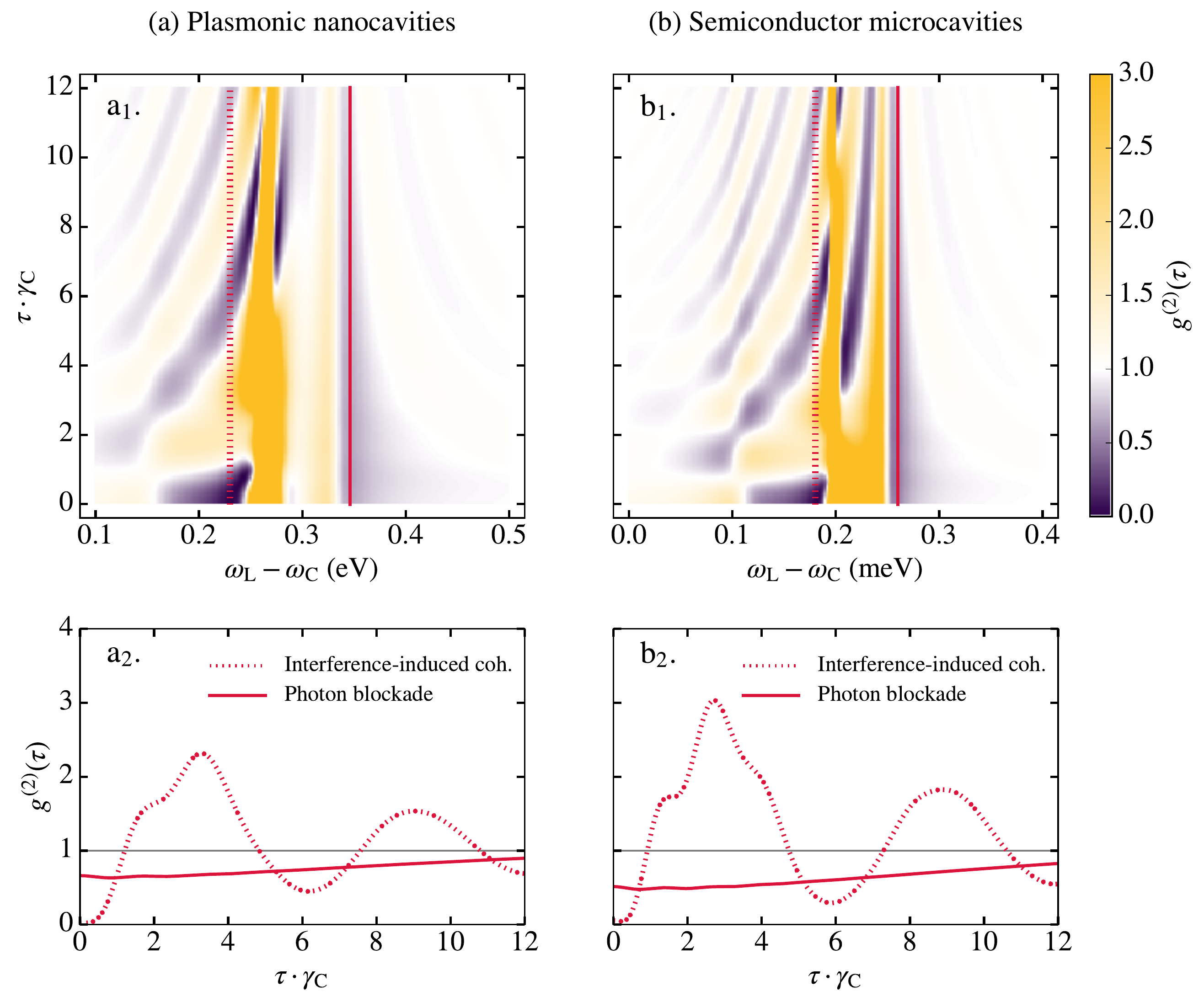}
    \caption{Correlation function $g^{(2)}(\tau)$ for an ensemble of $N=5$ quantum emitters interacting with either a nano- (a) or a micro- (b) cavity when $\Theta = 3\gamma_{\C}$. In top panels, $\gtau$ is plotted versus laser detuning $\omega_{\LL} - \omega_{\C}$ and time interval $\tau$ (in units of the cavity time $1/\gamma_{\C}$) for coupling strength $\lambda = 0.5 \gamma_{\C}$ in (a) and $\lambda = 0.8 \gamma_{\C}$ in (b). The cuts at $\tau =0$  correspond to the continuous pink lines depicted in \autoref{fig:det} (c). Bottom panels show cuts of the contour plots on top at laser detunings yielding two minima in the $\gz$ function. These specific configurations are indicated by red lines in the top panels and by red markers in \autoref{fig:det} (c).}
    \label{fig:tauDet}
\end{figure*}

Dotted lines show instead the evolution in $\tau$ for
configurations displaying interference-induced correlations. We
first observe that the degree of coherence at $\tau = 0$ reached
from this effect is stronger than the associated with the photon
blockade mechanism, although the couplings are smaller: for the
nanocavity, $\lambda / \gamma_{\C} = 0.2$, while for the
microcavity it takes the values $\lambda / \gamma_{\C} = 0.04$,
0.1 and 0.3 for $N=1$, 5 and 25 respectively. Oscillations in the
$\tau$-evolution of the function $\gtau$ are observed as $N$
increases, although  we still have antibunched light since $\gz <
\gtau$. When comparing nano- and microcavities, we observe that
oscillations in the latter are more pronounced. Moreover, note
again the difference in $\gamma_{\C}$, which translates into the
fact that the temporal evolution is significantly faster in the
plasmonic nanocavity (a direct consequence of the spectrally broad
character of photon correlations in the system).

A more general picture is shown in the bottom row of
\autoref{fig:tau}, where $\gtau$ is plotted as a function of the
laser detuning $\omega_{\LL} - \omega_0$ and the time delay $\tau$
(again in units of the cavity time $1/\gamma_{\C}$) for a
collection of $N=5$ emitters also interacting with either a nano-
(a) or a microcavity (b) for specific coupling strengths (see
figure caption). The particular values of the laser detuning
marked with vertical green lines (continuous for photon blockade
and dotted for interference-induced correlations) correspond to
the ones depicted in (a$_1$) and (b$_1$) for $N=5$. These contour
plots show that oscillatory patterns are also present for
configurations displaying super-Poissonian statistics at zero time
delay.

We have thus found that sub-Poissonian statistics is accompanied by antibunched light in the zero-detuning configurations explored in \autoref{fig:tau}.
Although $\gtau$ approaches one as the time delay increases, its
evolution is far from monotonous for interference-induced
correlations---they present an oscillatory pattern taking values
above and below one before reaching the coherent limit. This also
happens when detuning between the cavity frequency and the
emitters is introduced. An example is shown in the top row of
\autoref{fig:tauDet}, where the function $\gtau$ is plotted for a
particular coupling strength as a function of laser detuning
$\omega_{\LL} - \omega_{\C}$, and the time delay $\tau$ (in units
of $1/ \gamma_{\C}$) for $N=5$ quantum emitters interacting with
either a nano- (a) or a micro- (b) cavity. Here, we have
considered a detuning $\Theta = 3 \gamma_{\C}$, so these plots
corresponds to horizontal cuts in the panels of the third column of
\autoref{fig:det} (indicated by horizontal pink lines),
at $\lambda/\gamma_{\C} = 0.5$ for
plasmonic nanocavities and $\lambda/\gamma_{\C} = 0.8$ for
dielectric microcavities. In these
panels, for most laser detunings, the correlation function
develops an oscillatory pattern as $\tau$ increases for both sub-
and super-Poissonian statistics at $\tau=0$. Again, the close
similarity between the patterns for both cavities is remarkable
(once the delay time is expressed in units of $1/\gamma_{\C}$).

We observe that there exists a significant difference in the
temporal dependence of negative correlations also once detuning
between cavity and emitters is introduced. This is evident in the
bottom panels of \autoref{fig:tauDet}, where two specific values
of $\omega_{\LL} - \omega_{\C}$ are considered (indicated in
\autoref{fig:det} with magenta marks) in order to select
configurations that displays sub-Poissonian statistics due to
interference effects (dotted line) and photon blockade (continuous
line). These plots of $\gtau$ versus the time delay correspond to
vertical cuts in panels (a$_1$) and (b$_1$), see vertical red
lines. In the case of photon blockade, $\gtau$ approaches one
monotonically as the delay increases. In contrast, quantum
interference leads to an oscillatory pattern in correlations. Both
retain a temporal evolution similar to the one obtained at $\Theta
=0$ in \autoref{fig:tau}. Note that even the temporal slope and
pitch of oscillations remain the same. Thus, by detuning cavity
and emitters, the opportunity to obtain sub-Poissonian light
improves (as we have mentioned before, the parameter regions
widen) without altering qualitatively its evolution with time
delay between photon detections. Again, the phenomenology for
nano- and microcavities coincide, given that the values of laser
detuning and time delay, both normalized to the cavity losses, are
the same.

\section{Conclusions} \label{sec:Conclusions}

This work investigates the statistical properties of the light
generated by a collection of quantum emitters coupled to a single
electromagnetic mode. Theoretical computations based on an
effective Hamiltonian approach have been carried out to describe
the response of two different systems under low-intensity
coherent driving: plasmonic nanocavities and dielectric
microcavities. Special attention has focused on exploring the
impact that the distinct open/closed character of these two types
of cavities has on the scattered light.

For both cavity configurations, sub-Poissonian emission has been
observed not only at the single-emitter level, but also for
mesoscopic ensembles involving several tens of emitters. Our
results show that there are two different mechanisms that yield
significant negative correlations in the interaction between a
purely bosonic subsystem (cavity) and a quasi-bosonic one (emitter
ensemble): photon blockade and destructive interference. The
former takes place at high coupling strengths (comparable to or
larger than the cavity decay rate), while the latter becomes
relevant for weaker cavity-emitter interactions. Despite their
distinct open/closed character and the largely different physical
parameters describing nano- and microcavities, the photon
statistics phenomenology for both systems is remarkably similar
(once normalized to the cavity losses). This fact becomes clearer
through the exploration of cavity-emitter spectral detuning, which
enlarges the parameter range yielding antibunched light, and the
temporal evolution of correlations, which reveals the slow (fast)
fading of photon blockade (interference-induced) antibunching. Our findings may serve as
guidance for the optimization of quantum optical phenomena for
specific applications through the appropriate choice of material
parameters for their implementation.

%
%

\section*{Acknowledgments}

This work has been funded by the European Research Council under
Grant Agreements ERC-2011-AdG 290981 and ERC-2016-STG-714870, the EU Seventh Framework
Programme (FP7-PEOPLE-2013-CIG-630996 and
FP7-PEOPLE-2013-CIG-618229), and the Spanish MINECO under
contracts MAT2014-53432-C5-5-R and FIS2015-64951-R, as well as
through the ``Mar\'ia de Maeztu'' programme for Units of
Excellence in R\&D (MDM-2014-0377).


\begin{thebibliography}{81}%
    \makeatletter
    \providecommand \@ifxundefined [1]{%
        \@ifx{#1\undefined}
    }%
    \providecommand \@ifnum [1]{%
        \ifnum #1\expandafter \@firstoftwo
        \else \expandafter \@secondoftwo
        \fi
    }%
    \providecommand \@ifx [1]{%
        \ifx #1\expandafter \@firstoftwo
        \else \expandafter \@secondoftwo
        \fi
    }%
    \providecommand \natexlab [1]{#1}%
    \providecommand \enquote  [1]{``#1''}%
    \providecommand \bibnamefont  [1]{#1}%
    \providecommand \bibfnamefont [1]{#1}%
    \providecommand \citenamefont [1]{#1}%
    \providecommand \href@noop [0]{\@secondoftwo}%
    \providecommand \href [0]{\begingroup \@sanitize@url \@href}%
    \providecommand \@href[1]{\@@startlink{#1}\@@href}%
    \providecommand \@@href[1]{\endgroup#1\@@endlink}%
    \providecommand \@sanitize@url [0]{\catcode `\\12\catcode `\$12\catcode
        `\&12\catcode `\#12\catcode `\^12\catcode `\_12\catcode `\%12\relax}%
    \providecommand \@@startlink[1]{}%
    \providecommand \@@endlink[0]{}%
    \providecommand \url  [0]{\begingroup\@sanitize@url \@url }%
    \providecommand \@url [1]{\endgroup\@href {#1}{\urlprefix }}%
    \providecommand \urlprefix  [0]{URL }%
    \providecommand \Eprint [0]{\href }%
    \providecommand \doibase [0]{http://dx.doi.org/}%
    \providecommand \selectlanguage [0]{\@gobble}%
    \providecommand \bibinfo  [0]{\@secondoftwo}%
    \providecommand \bibfield  [0]{\@secondoftwo}%
    \providecommand \translation [1]{[#1]}%
    \providecommand \BibitemOpen [0]{}%
    \providecommand \bibitemStop [0]{}%
    \providecommand \bibitemNoStop [0]{.\EOS\space}%
    \providecommand \EOS [0]{\spacefactor3000\relax}%
    \providecommand \BibitemShut  [1]{\csname bibitem#1\endcsname}%
    \let\auto@bib@innerbib\@empty
    \bibitem [{\citenamefont {Kavokin}\ \emph {et~al.}(2007)\citenamefont
        {Kavokin}, \citenamefont {Baumberg}, \citenamefont {Malpuech},\ and\
        \citenamefont {Laussy}}]{MicrocavitiesBook}%
    \BibitemOpen
    \bibfield  {author} {\bibinfo {author} {\bibfnamefont {A.~V.}\ \bibnamefont
            {Kavokin}}, \bibinfo {author} {\bibfnamefont {J.~J.}\ \bibnamefont
            {Baumberg}}, \bibinfo {author} {\bibfnamefont {G.}~\bibnamefont {Malpuech}},
        \ and\ \bibinfo {author} {\bibfnamefont {F.~P.}\ \bibnamefont {Laussy}},\
    }\href@noop {} {\emph {\bibinfo {title} {Microcavities}}}\ (\bibinfo
    {publisher} {Oxford University Press},\ \bibinfo {address} {New York},\
    \bibinfo {year} {2007})\BibitemShut {NoStop}%
    \bibitem [{\citenamefont {Jaynes}\ and\ \citenamefont
        {Cummings}(1963)}]{JaynesCummings1963}%
    \BibitemOpen
    \bibfield  {author} {\bibinfo {author} {\bibfnamefont {E.~T.}\ \bibnamefont
            {Jaynes}}\ and\ \bibinfo {author} {\bibfnamefont {F.~W.}\ \bibnamefont
            {Cummings}},\ }\href@noop {} {\bibfield  {journal} {\bibinfo  {journal}
            {Proc. IEEE}\ }\textbf {\bibinfo {volume} {51}},\ \bibinfo {pages} {89}
        (\bibinfo {year} {1963})}\BibitemShut {NoStop}%
    \bibitem [{\citenamefont {Tavis}\ and\ \citenamefont
        {Cummings}(1968)}]{TavisCummings1968}%
    \BibitemOpen
    \bibfield  {author} {\bibinfo {author} {\bibfnamefont {M.}~\bibnamefont
            {Tavis}}\ and\ \bibinfo {author} {\bibfnamefont {F.~W.}\ \bibnamefont
            {Cummings}},\ }\href@noop {} {\bibfield  {journal} {\bibinfo  {journal}
            {Phys. Rev.}\ }\textbf {\bibinfo {volume} {170}},\ \bibinfo {pages} {379}
        (\bibinfo {year} {1968})}\BibitemShut {NoStop}%
    \bibitem [{\citenamefont {Breuer}\ and\ \citenamefont
        {Petruccione}(2002)}]{OpenBook}%
    \BibitemOpen
    \bibfield  {author} {\bibinfo {author} {\bibfnamefont {H.-P.}\ \bibnamefont
            {Breuer}}\ and\ \bibinfo {author} {\bibfnamefont {F.}~\bibnamefont
            {Petruccione}},\ }\href@noop {} {\emph {\bibinfo {title} {The theory of open
                quantum systems}}}\ (\bibinfo  {publisher} {Oxford University Press},\
    \bibinfo {address} {Oxford},\ \bibinfo {year} {2002})\BibitemShut {NoStop}%
    \bibitem [{\citenamefont {Michler}\ \emph {et~al.}(2000)\citenamefont
        {Michler}, \citenamefont {Kiraz}, \citenamefont {Zhang}, \citenamefont
        {Becher}, \citenamefont {Hu},\ and\ \citenamefont
        {Imamo$\breve{\g}$lu}}]{Lasing1}%
    \BibitemOpen
    \bibfield  {author} {\bibinfo {author} {\bibfnamefont {P.}~\bibnamefont
            {Michler}}, \bibinfo {author} {\bibfnamefont {A.}~\bibnamefont {Kiraz}},
        \bibinfo {author} {\bibfnamefont {L.}~\bibnamefont {Zhang}}, \bibinfo
        {author} {\bibfnamefont {C.}~\bibnamefont {Becher}}, \bibinfo {author}
        {\bibfnamefont {E.}~\bibnamefont {Hu}}, \ and\ \bibinfo {author}
        {\bibfnamefont {A.}~\bibnamefont {Imamo$\breve{\g}$lu}},\ }\href@noop {}
    {\bibfield  {journal} {\bibinfo  {journal} {Appl. Phys. Lett.}\ }\textbf
        {\bibinfo {volume} {77}},\ \bibinfo {pages} {184} (\bibinfo {year}
        {2000})}\BibitemShut {NoStop}%
    \bibitem [{\citenamefont {Reitzenstein}\ \emph {et~al.}(2006)\citenamefont
        {Reitzenstein}, \citenamefont {Bazhenov}, \citenamefont {Gorbunov},
        \citenamefont {Hofmann}, \citenamefont {M{\"u}nch}, \citenamefont
        {L{\"o}ffler}, \citenamefont {Kamp}, \citenamefont {Reithmaier},
        \citenamefont {Kulakovskii},\ and\ \citenamefont {Forchel}}]{Lasing2}%
    \BibitemOpen
    \bibfield  {author} {\bibinfo {author} {\bibfnamefont {S.}~\bibnamefont
            {Reitzenstein}}, \bibinfo {author} {\bibfnamefont {A.}~\bibnamefont
            {Bazhenov}}, \bibinfo {author} {\bibfnamefont {A.}~\bibnamefont {Gorbunov}},
        \bibinfo {author} {\bibfnamefont {C.}~\bibnamefont {Hofmann}}, \bibinfo
        {author} {\bibfnamefont {S.}~\bibnamefont {M{\"u}nch}}, \bibinfo {author}
        {\bibfnamefont {A.}~\bibnamefont {L{\"o}ffler}}, \bibinfo {author}
        {\bibfnamefont {M.}~\bibnamefont {Kamp}}, \bibinfo {author} {\bibfnamefont
            {J.~P.}\ \bibnamefont {Reithmaier}}, \bibinfo {author} {\bibfnamefont
            {V.~D.}\ \bibnamefont {Kulakovskii}}, \ and\ \bibinfo {author} {\bibfnamefont
            {A.}~\bibnamefont {Forchel}},\ }\href@noop {} {\bibfield  {journal} {\bibinfo
            {journal} {Appl. Phys. Lett.}\ }\textbf {\bibinfo {volume} {89}},\ \bibinfo
        {pages} {051107} (\bibinfo {year} {2006})}\BibitemShut {NoStop}%
    \bibitem [{\citenamefont {Gross}\ and\ \citenamefont
        {Haroche}(1982)}]{Haroche1982}%
    \BibitemOpen
    \bibfield  {author} {\bibinfo {author} {\bibfnamefont {M.}~\bibnamefont
            {Gross}}\ and\ \bibinfo {author} {\bibfnamefont {S.}~\bibnamefont
            {Haroche}},\ }\href@noop {} {\bibfield  {journal} {\bibinfo  {journal} {Phys.
                Rep.}\ }\textbf {\bibinfo {volume} {93}},\ \bibinfo {pages} {301} (\bibinfo
        {year} {1982})}\BibitemShut {NoStop}%
    \bibitem [{\citenamefont {Scheibner}\ \emph {et~al.}(2007)\citenamefont
        {Scheibner}, \citenamefont {Schmidt}, \citenamefont {Worschech},
        \citenamefont {Forchel}, \citenamefont {Bacher}, \citenamefont {Passow},\
        and\ \citenamefont {Hommel}}]{Hommel2007}%
    \BibitemOpen
    \bibfield  {author} {\bibinfo {author} {\bibfnamefont {M.}~\bibnamefont
            {Scheibner}}, \bibinfo {author} {\bibfnamefont {T.}~\bibnamefont {Schmidt}},
        \bibinfo {author} {\bibfnamefont {L.}~\bibnamefont {Worschech}}, \bibinfo
        {author} {\bibfnamefont {A.}~\bibnamefont {Forchel}}, \bibinfo {author}
        {\bibfnamefont {G.}~\bibnamefont {Bacher}}, \bibinfo {author} {\bibfnamefont
            {T.}~\bibnamefont {Passow}}, \ and\ \bibinfo {author} {\bibfnamefont
            {D.}~\bibnamefont {Hommel}},\ }\href@noop {} {\bibfield  {journal} {\bibinfo
            {journal} {Nat. Phys.}\ }\textbf {\bibinfo {volume} {3}},\ \bibinfo {pages}
        {106} (\bibinfo {year} {2007})}\BibitemShut {NoStop}%
    \bibitem [{\citenamefont {Faraon}\ \emph {et~al.}(2008)\citenamefont {Faraon},
        \citenamefont {Fushman}, \citenamefont {Englund}, \citenamefont {Stoltz},
        \citenamefont {Petroff},\ and\ \citenamefont
        {Vu$\check{\rm{c}}$kovi$\grave{\rm{c}}$}}]{VuckovicNat2008}%
    \BibitemOpen
    \bibfield  {author} {\bibinfo {author} {\bibfnamefont {A.}~\bibnamefont
            {Faraon}}, \bibinfo {author} {\bibfnamefont {I.}~\bibnamefont {Fushman}},
        \bibinfo {author} {\bibfnamefont {D.}~\bibnamefont {Englund}}, \bibinfo
        {author} {\bibfnamefont {N.}~\bibnamefont {Stoltz}}, \bibinfo {author}
        {\bibfnamefont {P.}~\bibnamefont {Petroff}}, \ and\ \bibinfo {author}
        {\bibfnamefont {J.}~\bibnamefont {Vu$\check{\rm{c}}$kovi$\grave{\rm{c}}$}},\
    }\href@noop {} {\bibfield  {journal} {\bibinfo  {journal} {Nat. Phys.}\
        }\textbf {\bibinfo {volume} {4}},\ \bibinfo {pages} {859} (\bibinfo {year}
        {2008})}\BibitemShut {NoStop}%
    \bibitem [{\citenamefont {Raimond}\ \emph {et~al.}(2001)\citenamefont
        {Raimond}, \citenamefont {Brune},\ and\ \citenamefont
        {Haroche}}]{Haroche2001}%
    \BibitemOpen
    \bibfield  {author} {\bibinfo {author} {\bibfnamefont {J.~M.}\ \bibnamefont
            {Raimond}}, \bibinfo {author} {\bibfnamefont {M.}~\bibnamefont {Brune}}, \
        and\ \bibinfo {author} {\bibfnamefont {S.}~\bibnamefont {Haroche}},\
    }\href@noop {} {\bibfield  {journal} {\bibinfo  {journal} {Rev. Mod. Phys.}\
        }\textbf {\bibinfo {volume} {73}},\ \bibinfo {pages} {565} (\bibinfo {year}
        {2001})}\BibitemShut {NoStop}%
    \bibitem [{\citenamefont {Pellizzari}\ \emph {et~al.}(1995)\citenamefont
        {Pellizzari}, \citenamefont {Gardiner}, \citenamefont {Cirac},\ and\
        \citenamefont {Zoller}}]{Zoller1995}%
    \BibitemOpen
    \bibfield  {author} {\bibinfo {author} {\bibfnamefont {T.}~\bibnamefont
            {Pellizzari}}, \bibinfo {author} {\bibfnamefont {S.~A.}\ \bibnamefont
            {Gardiner}}, \bibinfo {author} {\bibfnamefont {J.~I.}\ \bibnamefont {Cirac}},
        \ and\ \bibinfo {author} {\bibfnamefont {P.}~\bibnamefont {Zoller}},\
    }\href@noop {} {\bibfield  {journal} {\bibinfo  {journal} {Phys. Rev. Lett.}\
        }\textbf {\bibinfo {volume} {75}},\ \bibinfo {pages} {3788} (\bibinfo {year}
        {1995})}\BibitemShut {NoStop}%
    \bibitem [{\citenamefont {Imamo$\breve{\g}$lu}\ \emph
        {et~al.}(1999)\citenamefont {Imamo$\breve{\g}$lu}, \citenamefont {Awschalom},
        \citenamefont {Burkard}, \citenamefont {DiVincenzo}, \citenamefont {Loss},
        \citenamefont {Sherwin},\ and\ \citenamefont {Small}}]{Imamoglu1999}%
    \BibitemOpen
    \bibfield  {author} {\bibinfo {author} {\bibfnamefont {A.}~\bibnamefont
            {Imamo$\breve{\g}$lu}}, \bibinfo {author} {\bibfnamefont {D.~D.}\
            \bibnamefont {Awschalom}}, \bibinfo {author} {\bibfnamefont {G.}~\bibnamefont
            {Burkard}}, \bibinfo {author} {\bibfnamefont {D.~P.}\ \bibnamefont
            {DiVincenzo}}, \bibinfo {author} {\bibfnamefont {D.}~\bibnamefont {Loss}},
        \bibinfo {author} {\bibfnamefont {M.}~\bibnamefont {Sherwin}}, \ and\
        \bibinfo {author} {\bibfnamefont {A.}~\bibnamefont {Small}},\ }\href@noop {}
    {\bibfield  {journal} {\bibinfo  {journal} {Phys. Rev. Lett.}\ }\textbf
        {\bibinfo {volume} {83}},\ \bibinfo {pages} {4204} (\bibinfo {year}
        {1999})}\BibitemShut {NoStop}%
    \bibitem [{\citenamefont {Vahala}(2003)}]{Vahala2003}%
    \BibitemOpen
    \bibfield  {author} {\bibinfo {author} {\bibfnamefont {K.~J.}\ \bibnamefont
            {Vahala}},\ }\href@noop {} {\bibfield  {journal} {\bibinfo  {journal}
            {Nature}\ }\textbf {\bibinfo {volume} {424}},\ \bibinfo {pages} {839}
        (\bibinfo {year} {2003})}\BibitemShut {NoStop}%
    \bibitem [{\citenamefont {G{\'e}rard}\ \emph {et~al.}(1996)\citenamefont
        {G{\'e}rard}, \citenamefont {Barrier}, \citenamefont {Marzin}, \citenamefont
        {Kuszelewicz}, \citenamefont {Manin}, \citenamefont {Costard}, \citenamefont
        {Thierry-Mieg},\ and\ \citenamefont {Rivera}}]{pillarMicrocavity}%
    \BibitemOpen
    \bibfield  {author} {\bibinfo {author} {\bibfnamefont {J.~M.}\ \bibnamefont
            {G{\'e}rard}}, \bibinfo {author} {\bibfnamefont {D.}~\bibnamefont {Barrier}},
        \bibinfo {author} {\bibfnamefont {J.~Y.}\ \bibnamefont {Marzin}}, \bibinfo
        {author} {\bibfnamefont {R.}~\bibnamefont {Kuszelewicz}}, \bibinfo {author}
        {\bibfnamefont {L.}~\bibnamefont {Manin}}, \bibinfo {author} {\bibfnamefont
            {E.}~\bibnamefont {Costard}}, \bibinfo {author} {\bibfnamefont
            {V.}~\bibnamefont {Thierry-Mieg}}, \ and\ \bibinfo {author} {\bibfnamefont
            {T.}~\bibnamefont {Rivera}},\ }\href@noop {} {\bibfield  {journal} {\bibinfo
            {journal} {Appl. Phys. Lett.}\ }\textbf {\bibinfo {volume} {69}},\ \bibinfo
        {pages} {449} (\bibinfo {year} {1996})}\BibitemShut {NoStop}%
    \bibitem [{\citenamefont {Bennett}\ \emph {et~al.}(2005)\citenamefont
        {Bennett}, \citenamefont {Unitt}, \citenamefont {See}, \citenamefont
        {Shields}, \citenamefont {Atkinson}, \citenamefont {Cooper},\ and\
        \citenamefont {Ritchie}}]{braggMicrocavity}%
    \BibitemOpen
    \bibfield  {author} {\bibinfo {author} {\bibfnamefont {A.~J.}\ \bibnamefont
            {Bennett}}, \bibinfo {author} {\bibfnamefont {D.~C.}\ \bibnamefont {Unitt}},
        \bibinfo {author} {\bibfnamefont {P.}~\bibnamefont {See}}, \bibinfo {author}
        {\bibfnamefont {A.~J.}\ \bibnamefont {Shields}}, \bibinfo {author}
        {\bibfnamefont {P.}~\bibnamefont {Atkinson}}, \bibinfo {author}
        {\bibfnamefont {K.}~\bibnamefont {Cooper}}, \ and\ \bibinfo {author}
        {\bibfnamefont {D.~A.}\ \bibnamefont {Ritchie}},\ }\href@noop {} {\bibfield
        {journal} {\bibinfo  {journal} {Appl. Phys. Lett.}\ }\textbf {\bibinfo
            {volume} {86}},\ \bibinfo {pages} {181102} (\bibinfo {year}
        {2005})}\BibitemShut {NoStop}%
    \bibitem [{\citenamefont {Gayral}\ \emph {et~al.}(1999)\citenamefont {Gayral},
        \citenamefont {G{\'e}rard}, \citenamefont {Lema$\hat{ \mbox{\i}}$tre},
        \citenamefont {Dupuis}, \citenamefont {Manin},\ and\ \citenamefont
        {Pelouard}}]{whisper1}%
    \BibitemOpen
    \bibfield  {author} {\bibinfo {author} {\bibfnamefont {B.}~\bibnamefont
            {Gayral}}, \bibinfo {author} {\bibfnamefont {J.~M.}\ \bibnamefont
            {G{\'e}rard}}, \bibinfo {author} {\bibfnamefont {A.}~\bibnamefont {Lema$\hat{
                    \mbox{\i}}$tre}}, \bibinfo {author} {\bibfnamefont {C.}~\bibnamefont
            {Dupuis}}, \bibinfo {author} {\bibfnamefont {L.}~\bibnamefont {Manin}}, \
        and\ \bibinfo {author} {\bibfnamefont {J.~L.}\ \bibnamefont {Pelouard}},\
    }\href@noop {} {\bibfield  {journal} {\bibinfo  {journal} {Appl. Phys.
                Lett.}\ }\textbf {\bibinfo {volume} {75}},\ \bibinfo {pages} {1908} (\bibinfo
        {year} {1999})}\BibitemShut {NoStop}%
    \bibitem [{\citenamefont {Vernooy}\ \emph {et~al.}(1998)\citenamefont
        {Vernooy}, \citenamefont {Ilchenko}, \citenamefont {Mabuchi}, \citenamefont
        {Streed},\ and\ \citenamefont {Kimble}}]{whisper2}%
    \BibitemOpen
    \bibfield  {author} {\bibinfo {author} {\bibfnamefont {D.~W.}\ \bibnamefont
            {Vernooy}}, \bibinfo {author} {\bibfnamefont {V.~S.}\ \bibnamefont
            {Ilchenko}}, \bibinfo {author} {\bibfnamefont {H.}~\bibnamefont {Mabuchi}},
        \bibinfo {author} {\bibfnamefont {E.~W.}\ \bibnamefont {Streed}}, \ and\
        \bibinfo {author} {\bibfnamefont {H.~J.}\ \bibnamefont {Kimble}},\
    }\href@noop {} {\bibfield  {journal} {\bibinfo  {journal} {Opt. Lett.}\
        }\textbf {\bibinfo {volume} {23}},\ \bibinfo {pages} {247} (\bibinfo {year}
        {1998})}\BibitemShut {NoStop}%
    \bibitem [{\citenamefont {Srinivasan}\ \emph {et~al.}(2003)\citenamefont
        {Srinivasan}, \citenamefont {Barclay}, \citenamefont {Painter}, \citenamefont
        {Chen}, \citenamefont {Cho},\ and\ \citenamefont
        {Gmachl}}]{microPhotCrystal1}%
    \BibitemOpen
    \bibfield  {author} {\bibinfo {author} {\bibfnamefont {K.}~\bibnamefont
            {Srinivasan}}, \bibinfo {author} {\bibfnamefont {P.~E.}\ \bibnamefont
            {Barclay}}, \bibinfo {author} {\bibfnamefont {O.}~\bibnamefont {Painter}},
        \bibinfo {author} {\bibfnamefont {J.}~\bibnamefont {Chen}}, \bibinfo {author}
        {\bibfnamefont {A.~Y.}\ \bibnamefont {Cho}}, \ and\ \bibinfo {author}
        {\bibfnamefont {C.}~\bibnamefont {Gmachl}},\ }\href@noop {} {\bibfield
        {journal} {\bibinfo  {journal} {Appl. Phys. Lett.}\ }\textbf {\bibinfo
            {volume} {83}},\ \bibinfo {pages} {1915} (\bibinfo {year}
        {2003})}\BibitemShut {NoStop}%
    \bibitem [{\citenamefont {Vu$\check{\rm{c}}$kovi$\grave{\rm{c}}$}\ and\
        \citenamefont {Yoshihisa}(2003)}]{microPhotCrystal2}%
    \BibitemOpen
    \bibfield  {author} {\bibinfo {author} {\bibfnamefont {J.}~\bibnamefont
            {Vu$\check{\rm{c}}$kovi$\grave{\rm{c}}$}}\ and\ \bibinfo {author}
        {\bibfnamefont {Y.}~\bibnamefont {Yoshihisa}},\ }\href@noop {} {\bibfield
        {journal} {\bibinfo  {journal} {Appl. Phys. Lett.}\ }\textbf {\bibinfo
            {volume} {82}},\ \bibinfo {pages} {2374} (\bibinfo {year}
        {2003})}\BibitemShut {NoStop}%
    \bibitem [{\citenamefont {Akahane}\ \emph
        {et~al.}(2003{\natexlab{a}})\citenamefont {Akahane}, \citenamefont {Asano},
        \citenamefont {Song},\ and\ \citenamefont {Noda}}]{nanoPhotCrystal}%
    \BibitemOpen
    \bibfield  {author} {\bibinfo {author} {\bibfnamefont {Y.}~\bibnamefont
            {Akahane}}, \bibinfo {author} {\bibfnamefont {T.}~\bibnamefont {Asano}},
        \bibinfo {author} {\bibfnamefont {B.~S.}\ \bibnamefont {Song}}, \ and\
        \bibinfo {author} {\bibfnamefont {S.}~\bibnamefont {Noda}},\ }\href@noop {}
    {\bibfield  {journal} {\bibinfo  {journal} {Nature}\ }\textbf {\bibinfo
            {volume} {425}},\ \bibinfo {pages} {944} (\bibinfo {year}
        {2003}{\natexlab{a}})}\BibitemShut {NoStop}%
    \bibitem [{\citenamefont {Purcell}(1946)}]{Purcell1946}%
    \BibitemOpen
    \bibfield  {author} {\bibinfo {author} {\bibfnamefont {E.~M.}\ \bibnamefont
            {Purcell}},\ }\href@noop {} {\bibfield  {journal} {\bibinfo  {journal} {Phys.
                Rev.}\ }\textbf {\bibinfo {volume} {69}},\ \bibinfo {pages} {681} (\bibinfo
        {year} {1946})}\BibitemShut {NoStop}%
    \bibitem [{\citenamefont {Weisbuch}\ \emph {et~al.}(1992)\citenamefont
        {Weisbuch}, \citenamefont {Nishioka}, \citenamefont {Ishikawa},\ and\
        \citenamefont {Arakawa}}]{Arakawa1992}%
    \BibitemOpen
    \bibfield  {author} {\bibinfo {author} {\bibfnamefont {C.}~\bibnamefont
            {Weisbuch}}, \bibinfo {author} {\bibfnamefont {M.}~\bibnamefont {Nishioka}},
        \bibinfo {author} {\bibfnamefont {A.}~\bibnamefont {Ishikawa}}, \ and\
        \bibinfo {author} {\bibfnamefont {Y.}~\bibnamefont {Arakawa}},\ }\href@noop
    {} {\bibfield  {journal} {\bibinfo  {journal} {Phys. Rev. Lett.}\ }\textbf
        {\bibinfo {volume} {69}},\ \bibinfo {pages} {3314} (\bibinfo {year}
        {1992})}\BibitemShut {NoStop}%
    \bibitem [{\citenamefont {Houdre}\ \emph {et~al.}(1994)\citenamefont {Houdre},
        \citenamefont {Stanley}, \citenamefont {Oesterle}, \citenamefont {Ilegems},\
        and\ \citenamefont {Weisbuch}}]{Weisbuch1994}%
    \BibitemOpen
    \bibfield  {author} {\bibinfo {author} {\bibfnamefont {R.}~\bibnamefont
            {Houdre}}, \bibinfo {author} {\bibfnamefont {R.~P.}\ \bibnamefont {Stanley}},
        \bibinfo {author} {\bibfnamefont {U.}~\bibnamefont {Oesterle}}, \bibinfo
        {author} {\bibfnamefont {M.}~\bibnamefont {Ilegems}}, \ and\ \bibinfo
        {author} {\bibfnamefont {C.}~\bibnamefont {Weisbuch}},\ }\href@noop {}
    {\bibfield  {journal} {\bibinfo  {journal} {Phys. Rev. B}\ }\textbf {\bibinfo
            {volume} {49}},\ \bibinfo {pages} {16761} (\bibinfo {year}
        {1994})}\BibitemShut {NoStop}%
    \bibitem [{\citenamefont {Bellessa}\ \emph {et~al.}(2004)\citenamefont
        {Bellessa}, \citenamefont {Bonnand}, \citenamefont {Plenet},\ and\
        \citenamefont {Mugnier}}]{Bellesa2004}%
    \BibitemOpen
    \bibfield  {author} {\bibinfo {author} {\bibfnamefont {J.}~\bibnamefont
            {Bellessa}}, \bibinfo {author} {\bibfnamefont {C.}~\bibnamefont {Bonnand}},
        \bibinfo {author} {\bibfnamefont {J.~C.}\ \bibnamefont {Plenet}}, \ and\
        \bibinfo {author} {\bibfnamefont {J.}~\bibnamefont {Mugnier}},\ }\href@noop
    {} {\bibfield  {journal} {\bibinfo  {journal} {Phys. Rev. Lett.}\ }\textbf
        {\bibinfo {volume} {93}},\ \bibinfo {pages} {036404} (\bibinfo {year}
        {2004})}\BibitemShut {NoStop}%
    \bibitem [{\citenamefont {Schwartz}\ \emph {et~al.}(2011)\citenamefont
        {Schwartz}, \citenamefont {Hutchison}, \citenamefont {Genet},\ and\
        \citenamefont {Ebbesen}}]{Ebbesen2011}%
    \BibitemOpen
    \bibfield  {author} {\bibinfo {author} {\bibfnamefont {T.}~\bibnamefont
            {Schwartz}}, \bibinfo {author} {\bibfnamefont {J.~A.}\ \bibnamefont
            {Hutchison}}, \bibinfo {author} {\bibfnamefont {C.}~\bibnamefont {Genet}}, \
        and\ \bibinfo {author} {\bibfnamefont {T.~W.}\ \bibnamefont {Ebbesen}},\
    }\href@noop {} {\bibfield  {journal} {\bibinfo  {journal} {Phys. Rev. Lett.}\
        }\textbf {\bibinfo {volume} {106}},\ \bibinfo {pages} {196405} (\bibinfo
        {year} {2011})}\BibitemShut {NoStop}%
    \bibitem [{\citenamefont {McKeever}\ \emph {et~al.}(2003)\citenamefont
        {McKeever}, \citenamefont {Boca}, \citenamefont {Boozer}, \citenamefont
        {Buck},\ and\ \citenamefont {Kimble}}]{Kimble2003}%
    \BibitemOpen
    \bibfield  {author} {\bibinfo {author} {\bibfnamefont {J.}~\bibnamefont
            {McKeever}}, \bibinfo {author} {\bibfnamefont {A.}~\bibnamefont {Boca}},
        \bibinfo {author} {\bibfnamefont {A.~D.}\ \bibnamefont {Boozer}}, \bibinfo
        {author} {\bibfnamefont {J.~R.}\ \bibnamefont {Buck}}, \ and\ \bibinfo
        {author} {\bibfnamefont {H.~J.}\ \bibnamefont {Kimble}},\ }\href@noop {}
    {\bibfield  {journal} {\bibinfo  {journal} {Nature}\ }\textbf {\bibinfo
            {volume} {425}},\ \bibinfo {pages} {268} (\bibinfo {year}
        {2003})}\BibitemShut {NoStop}%
    \bibitem [{\citenamefont {Reithmaier}\ \emph {et~al.}(2004)\citenamefont
        {Reithmaier}, \citenamefont {G.~S$\k{e}$k}, \citenamefont {Hofmann},
        \citenamefont {Kuhn}, \citenamefont {Reitzenstein}, \citenamefont {Keldysh},
        \citenamefont {Kulakovskii}, \citenamefont {Reinecke},\ and\ \citenamefont
        {Forchel}}]{Forchel2004}%
    \BibitemOpen
    \bibfield  {author} {\bibinfo {author} {\bibfnamefont {J.~P.}\ \bibnamefont
            {Reithmaier}}, \bibinfo {author} {\bibfnamefont {A.~L.}\ \bibnamefont
            {G.~S$\k{e}$k}}, \bibinfo {author} {\bibfnamefont {C.}~\bibnamefont
            {Hofmann}}, \bibinfo {author} {\bibfnamefont {S.}~\bibnamefont {Kuhn}},
        \bibinfo {author} {\bibfnamefont {S.}~\bibnamefont {Reitzenstein}}, \bibinfo
        {author} {\bibfnamefont {L.~V.}\ \bibnamefont {Keldysh}}, \bibinfo {author}
        {\bibfnamefont {V.~D.}\ \bibnamefont {Kulakovskii}}, \bibinfo {author}
        {\bibfnamefont {T.~L.}\ \bibnamefont {Reinecke}}, \ and\ \bibinfo {author}
        {\bibfnamefont {A.}~\bibnamefont {Forchel}},\ }\href@noop {} {\bibfield
        {journal} {\bibinfo  {journal} {Nature}\ }\textbf {\bibinfo {volume} {432}},\
        \bibinfo {pages} {197} (\bibinfo {year} {2004})}\BibitemShut {NoStop}%
    \bibitem [{\citenamefont {Yoshie}\ \emph {et~al.}(2004)\citenamefont {Yoshie},
        \citenamefont {Scherer}, \citenamefont {Hendrickson}, \citenamefont
        {Khitrova}, \citenamefont {Gibbs}, \citenamefont {Rupper}, \citenamefont
        {Ell}, \citenamefont {Shchekin},\ and\ \citenamefont {Deppe}}]{Yoshie2004}%
    \BibitemOpen
    \bibfield  {author} {\bibinfo {author} {\bibfnamefont {T.}~\bibnamefont
            {Yoshie}}, \bibinfo {author} {\bibfnamefont {A.}~\bibnamefont {Scherer}},
        \bibinfo {author} {\bibfnamefont {J.}~\bibnamefont {Hendrickson}}, \bibinfo
        {author} {\bibfnamefont {G.}~\bibnamefont {Khitrova}}, \bibinfo {author}
        {\bibfnamefont {H.~M.}\ \bibnamefont {Gibbs}}, \bibinfo {author}
        {\bibfnamefont {G.}~\bibnamefont {Rupper}}, \bibinfo {author} {\bibfnamefont
            {C.}~\bibnamefont {Ell}}, \bibinfo {author} {\bibfnamefont {O.~B.}\
            \bibnamefont {Shchekin}}, \ and\ \bibinfo {author} {\bibfnamefont {D.~G.}\
            \bibnamefont {Deppe}},\ }\href@noop {} {\bibfield  {journal} {\bibinfo
            {journal} {Nature}\ }\textbf {\bibinfo {volume} {432}},\ \bibinfo {pages}
        {200} (\bibinfo {year} {2004})}\BibitemShut {NoStop}%
    \bibitem [{\citenamefont {Chikkaraddy}\ \emph {et~al.}(2016)\citenamefont
        {Chikkaraddy}, \citenamefont {Nijs}, \citenamefont {Benz}, \citenamefont
        {Barrow}, \citenamefont {Scherman}, \citenamefont {Rosta}, \citenamefont
        {Demetriadou}, \citenamefont {Fox}, \citenamefont {Hess},\ and\ \citenamefont
        {Baumberg}}]{Baumberg2016}%
    \BibitemOpen
    \bibfield  {author} {\bibinfo {author} {\bibfnamefont {R.}~\bibnamefont
            {Chikkaraddy}}, \bibinfo {author} {\bibfnamefont {B.~D.}\ \bibnamefont
            {Nijs}}, \bibinfo {author} {\bibfnamefont {F.}~\bibnamefont {Benz}}, \bibinfo
        {author} {\bibfnamefont {S.~J.}\ \bibnamefont {Barrow}}, \bibinfo {author}
        {\bibfnamefont {O.~A.}\ \bibnamefont {Scherman}}, \bibinfo {author}
        {\bibfnamefont {E.}~\bibnamefont {Rosta}}, \bibinfo {author} {\bibfnamefont
            {A.}~\bibnamefont {Demetriadou}}, \bibinfo {author} {\bibfnamefont
            {P.}~\bibnamefont {Fox}}, \bibinfo {author} {\bibfnamefont {O.}~\bibnamefont
            {Hess}}, \ and\ \bibinfo {author} {\bibfnamefont {J.}~\bibnamefont
            {Baumberg}},\ }\href@noop {} {\bibfield  {journal} {\bibinfo  {journal}
            {Nature}\ }\textbf {\bibinfo {volume} {535}},\ \bibinfo {pages} {127}
        (\bibinfo {year} {2016})}\BibitemShut {NoStop}%
    \bibitem [{\citenamefont {Santhosh}\ \emph {et~al.}(2016)\citenamefont
        {Santhosh}, \citenamefont {Bitton}, \citenamefont {Chuntonov},\ and\
        \citenamefont {Haran}}]{Santhosh2016}%
    \BibitemOpen
    \bibfield  {author} {\bibinfo {author} {\bibfnamefont {K.}~\bibnamefont
            {Santhosh}}, \bibinfo {author} {\bibfnamefont {O.}~\bibnamefont {Bitton}},
        \bibinfo {author} {\bibfnamefont {L.}~\bibnamefont {Chuntonov}}, \ and\
        \bibinfo {author} {\bibfnamefont {G.}~\bibnamefont {Haran}},\ }\href@noop {}
    {\bibfield  {journal} {\bibinfo  {journal} {Nat. Comm.}\ }\textbf {\bibinfo
            {volume} {7}},\ \bibinfo {pages} {1182} (\bibinfo {year} {2016})}\BibitemShut
    {NoStop}%
    \bibitem [{\citenamefont {Dicke}(1954)}]{Dicke1954}%
    \BibitemOpen
    \bibfield  {author} {\bibinfo {author} {\bibfnamefont {R.~H.}\ \bibnamefont
            {Dicke}},\ }\href@noop {} {\bibfield  {journal} {\bibinfo  {journal} {Phys.
                Rev.}\ }\textbf {\bibinfo {volume} {93}},\ \bibinfo {pages} {99} (\bibinfo
        {year} {1954})}\BibitemShut {NoStop}%
    \bibitem [{\citenamefont {Delga}\ \emph {et~al.}(2014)\citenamefont {Delga},
        \citenamefont {Feist}, \citenamefont {Bravo-Abad},\ and\ \citenamefont
        {Garc{\'i}a-Vidal}}]{Delga2014}%
    \BibitemOpen
    \bibfield  {author} {\bibinfo {author} {\bibfnamefont {A.}~\bibnamefont
            {Delga}}, \bibinfo {author} {\bibfnamefont {J.}~\bibnamefont {Feist}},
        \bibinfo {author} {\bibfnamefont {J.}~\bibnamefont {Bravo-Abad}}, \ and\
        \bibinfo {author} {\bibfnamefont {F.~J.}\ \bibnamefont {Garc{\'i}a-Vidal}},\
    }\href@noop {} {\bibfield  {journal} {\bibinfo  {journal} {Phys. Rev. Lett.}\
        }\textbf {\bibinfo {volume} {112}},\ \bibinfo {pages} {253601} (\bibinfo
        {year} {2014})}\BibitemShut {NoStop}%
    \bibitem [{\citenamefont {Gonz{\'a}lez-Tudela}\ \emph
        {et~al.}(2013)\citenamefont {Gonz{\'a}lez-Tudela}, \citenamefont {Huidobro},
        \citenamefont {Mart{\'i}n-Moreno}, \citenamefont {Tejedor},\ and\
        \citenamefont {Garc{\'i}a-Vidal}}]{GonzalezTudela2013}%
    \BibitemOpen
    \bibfield  {author} {\bibinfo {author} {\bibfnamefont {A.}~\bibnamefont
            {Gonz{\'a}lez-Tudela}}, \bibinfo {author} {\bibfnamefont {P.~A.}\
            \bibnamefont {Huidobro}}, \bibinfo {author} {\bibfnamefont {L.}~\bibnamefont
            {Mart{\'i}n-Moreno}}, \bibinfo {author} {\bibfnamefont {C.}~\bibnamefont
            {Tejedor}}, \ and\ \bibinfo {author} {\bibfnamefont {F.~J.}\ \bibnamefont
            {Garc{\'i}a-Vidal}},\ }\href@noop {} {\bibfield  {journal} {\bibinfo
            {journal} {Phys. Rev. Lett.}\ }\textbf {\bibinfo {volume} {110}},\ \bibinfo
        {pages} {126801} (\bibinfo {year} {2013})}\BibitemShut {NoStop}%
    \bibitem [{\citenamefont {Shammah}\ \emph {et~al.}(2017)\citenamefont
        {Shammah}, \citenamefont {Lambert}, \citenamefont {Nori},\ and\ \citenamefont
        {DeLiberato}}]{DeLiberato2017}%
    \BibitemOpen
    \bibfield  {author} {\bibinfo {author} {\bibfnamefont {N.}~\bibnamefont
            {Shammah}}, \bibinfo {author} {\bibfnamefont {N.}~\bibnamefont {Lambert}},
        \bibinfo {author} {\bibfnamefont {F.}~\bibnamefont {Nori}}, \ and\ \bibinfo
        {author} {\bibfnamefont {S.}~\bibnamefont {DeLiberato}},\ }\href@noop {}
    {\bibfield  {journal} {\bibinfo  {journal} {Phys. Rev. A}\ }\textbf {\bibinfo
            {volume} {96}},\ \bibinfo {pages} {023863} (\bibinfo {year}
        {2017})}\BibitemShut {NoStop}%
    \bibitem [{\citenamefont {Alpeggiani}\ \emph {et~al.}(2016)\citenamefont
        {Alpeggiani}, \citenamefont {D'Agostino}, \citenamefont {Sanvitto},\ and\
        \citenamefont {Gerace}}]{Alpeggiani2016}%
    \BibitemOpen
    \bibfield  {author} {\bibinfo {author} {\bibfnamefont {F.}~\bibnamefont
            {Alpeggiani}}, \bibinfo {author} {\bibfnamefont {S.}~\bibnamefont
            {D'Agostino}}, \bibinfo {author} {\bibfnamefont {D.}~\bibnamefont
            {Sanvitto}}, \ and\ \bibinfo {author} {\bibfnamefont {D.}~\bibnamefont
            {Gerace}},\ }\href@noop {} {\bibfield  {journal} {\bibinfo  {journal} {Sci.
                Rep.}\ }\textbf {\bibinfo {volume} {6}},\ \bibinfo {pages} {34772} (\bibinfo
        {year} {2016})}\BibitemShut {NoStop}%
    \bibitem [{\citenamefont {S{\'a}ez-Bl{\'a}zquez}\ \emph
        {et~al.}(2017)\citenamefont {S{\'a}ez-Bl{\'a}zquez}, \citenamefont {Feist},
        \citenamefont {Fern{\'a}ndez-Dom{\'i}nguez},\ and\ \citenamefont
        {Garc{\'i}a-Vidal}}]{Rocio2017}%
    \BibitemOpen
    \bibfield  {author} {\bibinfo {author} {\bibfnamefont {R.}~\bibnamefont
            {S{\'a}ez-Bl{\'a}zquez}}, \bibinfo {author} {\bibfnamefont {J.}~\bibnamefont
            {Feist}}, \bibinfo {author} {\bibfnamefont {A.~I.}\ \bibnamefont
            {Fern{\'a}ndez-Dom{\'i}nguez}}, \ and\ \bibinfo {author} {\bibfnamefont
            {F.~J.}\ \bibnamefont {Garc{\'i}a-Vidal}},\ }\href@noop {} {\bibfield
        {journal} {\bibinfo  {journal} {Optica}\ }\textbf {\bibinfo {volume} {4}},\
        \bibinfo {pages} {1363} (\bibinfo {year} {2017})}\BibitemShut {NoStop}%
    \bibitem [{\citenamefont {Kimble}\ \emph {et~al.}(1977)\citenamefont {Kimble},
        \citenamefont {Dagenais},\ and\ \citenamefont {Mandel}}]{KimbleMandel1977}%
    \BibitemOpen
    \bibfield  {author} {\bibinfo {author} {\bibfnamefont {H.~J.}\ \bibnamefont
            {Kimble}}, \bibinfo {author} {\bibfnamefont {M.}~\bibnamefont {Dagenais}}, \
        and\ \bibinfo {author} {\bibfnamefont {L.}~\bibnamefont {Mandel}},\
    }\href@noop {} {\bibfield  {journal} {\bibinfo  {journal} {Phys. Rev. Lett.}\
        }\textbf {\bibinfo {volume} {39}},\ \bibinfo {pages} {691} (\bibinfo {year}
        {1977})}\BibitemShut {NoStop}%
    \bibitem [{\citenamefont {Carre{\~n}o}\ \emph {et~al.}(2016)\citenamefont
        {Carre{\~n}o}, \citenamefont {Casalengua}, \citenamefont {Valle},\ and\
        \citenamefont {Laussy}}]{Laussy}%
    \BibitemOpen
    \bibfield  {author} {\bibinfo {author} {\bibfnamefont {J.~C.~L.}\
            \bibnamefont {Carre{\~n}o}}, \bibinfo {author} {\bibfnamefont {E.~Z.}\
            \bibnamefont {Casalengua}}, \bibinfo {author} {\bibfnamefont {E.~D.}\
            \bibnamefont {Valle}}, \ and\ \bibinfo {author} {\bibfnamefont {F.~P.}\
            \bibnamefont {Laussy}},\ }\href@noop {} {\bibfield  {journal} {\bibinfo
            {journal} {arXiv:1610.06126}\ } (\bibinfo {year} {2016})}\BibitemShut
    {NoStop}%
    \bibitem [{\citenamefont {Bennett}\ \emph {et~al.}(1992)\citenamefont
        {Bennett}, \citenamefont {Brassard},\ and\ \citenamefont
        {Ekert}}]{quantumCryptography}%
    \BibitemOpen
    \bibfield  {author} {\bibinfo {author} {\bibfnamefont {C.~H.}\ \bibnamefont
            {Bennett}}, \bibinfo {author} {\bibfnamefont {G.}~\bibnamefont {Brassard}}, \
        and\ \bibinfo {author} {\bibfnamefont {A.~K.}\ \bibnamefont {Ekert}},\
    }\href@noop {} {\bibfield  {journal} {\bibinfo  {journal} {Sci. Am.}\
        }\textbf {\bibinfo {volume} {267}},\ \bibinfo {pages} {50} (\bibinfo {year}
        {1992})}\BibitemShut {NoStop}%
    \bibitem [{\citenamefont {Degen}\ \emph {et~al.}(2017)\citenamefont {Degen},
        \citenamefont {Reinhard},\ and\ \citenamefont {Cappellaro}}]{quantumSensing}%
    \BibitemOpen
    \bibfield  {author} {\bibinfo {author} {\bibfnamefont {C.~L.}\ \bibnamefont
            {Degen}}, \bibinfo {author} {\bibfnamefont {F.}~\bibnamefont {Reinhard}}, \
        and\ \bibinfo {author} {\bibfnamefont {P.}~\bibnamefont {Cappellaro}},\
    }\href@noop {} {\bibfield  {journal} {\bibinfo  {journal} {Rev. Mod. Phys.}\
        }\textbf {\bibinfo {volume} {89}},\ \bibinfo {pages} {035002} (\bibinfo
        {year} {2017})}\BibitemShut {NoStop}%
    \bibitem [{\citenamefont {Giovannetti}\ \emph {et~al.}(2006)\citenamefont
        {Giovannetti}, \citenamefont {Lloyd},\ and\ \citenamefont
        {Maccone}}]{quantumMetrology}%
    \BibitemOpen
    \bibfield  {author} {\bibinfo {author} {\bibfnamefont {V.}~\bibnamefont
            {Giovannetti}}, \bibinfo {author} {\bibfnamefont {S.}~\bibnamefont {Lloyd}},
        \ and\ \bibinfo {author} {\bibfnamefont {L.}~\bibnamefont {Maccone}},\
    }\href@noop {} {\bibfield  {journal} {\bibinfo  {journal} {Phys. Rev. Lett.}\
        }\textbf {\bibinfo {volume} {96}},\ \bibinfo {pages} {010401} (\bibinfo
        {year} {2006})}\BibitemShut {NoStop}%
    \bibitem [{\citenamefont {Kimble}(2008)}]{Kimble2008}%
    \BibitemOpen
    \bibfield  {author} {\bibinfo {author} {\bibfnamefont {H.~J.}\ \bibnamefont
            {Kimble}},\ }\href@noop {} {\bibfield  {journal} {\bibinfo  {journal}
            {Nature}\ }\textbf {\bibinfo {volume} {453}},\ \bibinfo {pages} {1023}
        (\bibinfo {year} {2008})}\BibitemShut {NoStop}%
    \bibitem [{\citenamefont {Lounis}\ and\ \citenamefont
        {Orrit}(2005)}]{SinglePhotons2005}%
    \BibitemOpen
    \bibfield  {author} {\bibinfo {author} {\bibfnamefont {B.}~\bibnamefont
            {Lounis}}\ and\ \bibinfo {author} {\bibfnamefont {M.}~\bibnamefont {Orrit}},\
    }\href@noop {} {\bibfield  {journal} {\bibinfo  {journal} {Rep. Prog. Phys.}\
        }\textbf {\bibinfo {volume} {68}},\ \bibinfo {pages} {1129} (\bibinfo {year}
        {2005})}\BibitemShut {NoStop}%
    \bibitem [{\citenamefont {Muller}\ \emph {et~al.}(1997)\citenamefont {Muller},
        \citenamefont {Herzog}, \citenamefont {Huttner}, \citenamefont {Tittel},
        \citenamefont {Zbinden},\ and\ \citenamefont {Gisin}}]{faintPulses}%
    \BibitemOpen
    \bibfield  {author} {\bibinfo {author} {\bibfnamefont {A.}~\bibnamefont
            {Muller}}, \bibinfo {author} {\bibfnamefont {T.}~\bibnamefont {Herzog}},
        \bibinfo {author} {\bibfnamefont {B.}~\bibnamefont {Huttner}}, \bibinfo
        {author} {\bibfnamefont {W.}~\bibnamefont {Tittel}}, \bibinfo {author}
        {\bibfnamefont {H.}~\bibnamefont {Zbinden}}, \ and\ \bibinfo {author}
        {\bibfnamefont {N.}~\bibnamefont {Gisin}},\ }\href@noop {} {\  (\bibinfo
        {year} {1997})}\BibitemShut {NoStop}%
    \bibitem [{\citenamefont {Kuhn}\ \emph {et~al.}(2002)\citenamefont {Kuhn},
        \citenamefont {Hennrich},\ and\ \citenamefont {Rempe}}]{singleAtom}%
    \BibitemOpen
    \bibfield  {author} {\bibinfo {author} {\bibfnamefont {A.}~\bibnamefont
            {Kuhn}}, \bibinfo {author} {\bibfnamefont {M.}~\bibnamefont {Hennrich}}, \
        and\ \bibinfo {author} {\bibfnamefont {G.}~\bibnamefont {Rempe}},\
    }\href@noop {} {\bibfield  {journal} {\bibinfo  {journal} {Phys. Rev. Lett.}\
        }\textbf {\bibinfo {volume} {89}},\ \bibinfo {pages} {067901} (\bibinfo
        {year} {2002})}\BibitemShut {NoStop}%
    \bibitem [{\citenamefont {Lounis}\ and\ \citenamefont
        {Moerner}(2000)}]{singleMolecule}%
    \BibitemOpen
    \bibfield  {author} {\bibinfo {author} {\bibfnamefont {B.}~\bibnamefont
            {Lounis}}\ and\ \bibinfo {author} {\bibfnamefont {W.~E.}\ \bibnamefont
            {Moerner}},\ }\href@noop {} {\  (\bibinfo {year} {2000})}\BibitemShut
    {NoStop}%
    \bibitem [{\citenamefont {Aharonovich}\ \emph {et~al.}(2011)\citenamefont
        {Aharonovich}, \citenamefont {Castelletto}, \citenamefont {Simpson},
        \citenamefont {Su}, \citenamefont {Greentree},\ and\ \citenamefont
        {Prawer}}]{diamond}%
    \BibitemOpen
    \bibfield  {author} {\bibinfo {author} {\bibfnamefont {I.}~\bibnamefont
            {Aharonovich}}, \bibinfo {author} {\bibfnamefont {S.}~\bibnamefont
            {Castelletto}}, \bibinfo {author} {\bibfnamefont {D.~A.}\ \bibnamefont
            {Simpson}}, \bibinfo {author} {\bibfnamefont {C.-H.}\ \bibnamefont {Su}},
        \bibinfo {author} {\bibfnamefont {A.~D.}\ \bibnamefont {Greentree}}, \ and\
        \bibinfo {author} {\bibfnamefont {S.}~\bibnamefont {Prawer}},\ }\href@noop {}
    {\bibfield  {journal} {\bibinfo  {journal} {Rep. Progr. Phys.}\ }\textbf
        {\bibinfo {volume} {74}},\ \bibinfo {pages} {076501} (\bibinfo {year}
        {2011})}\BibitemShut {NoStop}%
    \bibitem [{\citenamefont {Brokmann}\ \emph {et~al.}(2004)\citenamefont
        {Brokmann}, \citenamefont {Giacobino}, \citenamefont {Dahana},\ and\
        \citenamefont {Hermier}}]{nanocrystal}%
    \BibitemOpen
    \bibfield  {author} {\bibinfo {author} {\bibfnamefont {X.}~\bibnamefont
            {Brokmann}}, \bibinfo {author} {\bibfnamefont {E.}~\bibnamefont {Giacobino}},
        \bibinfo {author} {\bibfnamefont {M.}~\bibnamefont {Dahana}}, \ and\ \bibinfo
        {author} {\bibfnamefont {J.~P.}\ \bibnamefont {Hermier}},\ }\href@noop {}
    {\bibfield  {journal} {\bibinfo  {journal} {Appl. Phys. Lett.}\ }\textbf
        {\bibinfo {volume} {85}},\ \bibinfo {pages} {712} (\bibinfo {year}
        {2004})}\BibitemShut {NoStop}%
    \bibitem [{\citenamefont {Hong}\ and\ \citenamefont
        {Mandel}(1986)}]{parametricDownconversion}%
    \BibitemOpen
    \bibfield  {author} {\bibinfo {author} {\bibfnamefont {C.~K.}\ \bibnamefont
            {Hong}}\ and\ \bibinfo {author} {\bibfnamefont {L.}~\bibnamefont {Mandel}},\
    }\href@noop {} {\bibfield  {journal} {\bibinfo  {journal} {Phys. Rev. Lett.}\
        }\textbf {\bibinfo {volume} {56}},\ \bibinfo {pages} {58} (\bibinfo {year}
        {1986})}\BibitemShut {NoStop}%
    \bibitem [{\citenamefont {Rempe}\ \emph {et~al.}(1991)\citenamefont {Rempe},
        \citenamefont {Thompson}, \citenamefont {Brecha}, \citenamefont {Lee},\ and\
        \citenamefont {Kimble}}]{Kimble1991}%
    \BibitemOpen
    \bibfield  {author} {\bibinfo {author} {\bibfnamefont {G.}~\bibnamefont
            {Rempe}}, \bibinfo {author} {\bibfnamefont {R.~J.}\ \bibnamefont {Thompson}},
        \bibinfo {author} {\bibfnamefont {R.~J.}\ \bibnamefont {Brecha}}, \bibinfo
        {author} {\bibfnamefont {W.~D.}\ \bibnamefont {Lee}}, \ and\ \bibinfo
        {author} {\bibfnamefont {H.~J.}\ \bibnamefont {Kimble}},\ }\href@noop {}
    {\bibfield  {journal} {\bibinfo  {journal} {Phys. Rev. Lett.}\ }\textbf
        {\bibinfo {volume} {67}},\ \bibinfo {pages} {1727} (\bibinfo {year}
        {1991})}\BibitemShut {NoStop}%
    \bibitem [{\citenamefont {Carmichael}\ \emph {et~al.}(1991)\citenamefont
        {Carmichael}, \citenamefont {Brecha},\ and\ \citenamefont
        {Rice}}]{Carmichael1991}%
    \BibitemOpen
    \bibfield  {author} {\bibinfo {author} {\bibfnamefont {H.~J.}\ \bibnamefont
            {Carmichael}}, \bibinfo {author} {\bibfnamefont {R.~J.}\ \bibnamefont
            {Brecha}}, \ and\ \bibinfo {author} {\bibfnamefont {P.~R.}\ \bibnamefont
            {Rice}},\ }\href@noop {} {\bibfield  {journal} {\bibinfo  {journal} {Optics
                Communications}\ }\textbf {\bibinfo {volume} {82}},\ \bibinfo {pages} {73 }
        (\bibinfo {year} {1991})}\BibitemShut {NoStop}%
    \bibitem [{\citenamefont {Foster}\ \emph {et~al.}(2000)\citenamefont {Foster},
        \citenamefont {Mielke},\ and\ \citenamefont {Orozco}}]{Foster2000}%
    \BibitemOpen
    \bibfield  {author} {\bibinfo {author} {\bibfnamefont {G.~T.}\ \bibnamefont
            {Foster}}, \bibinfo {author} {\bibfnamefont {S.~L.}\ \bibnamefont {Mielke}},
        \ and\ \bibinfo {author} {\bibfnamefont {L.~A.}\ \bibnamefont {Orozco}},\
    }\href@noop {} {\bibfield  {journal} {\bibinfo  {journal} {Phys. Rev. A}\
        }\textbf {\bibinfo {volume} {61}},\ \bibinfo {pages} {053821} (\bibinfo
        {year} {2000})}\BibitemShut {NoStop}%
    \bibitem [{\citenamefont {Auff{\'e}ves}\ \emph {et~al.}(2011)\citenamefont
        {Auff{\'e}ves}, \citenamefont {Gerace}, \citenamefont {Portolan},
        \citenamefont {Drezet},\ and\ \citenamefont {Santos}}]{Auffeves2011}%
    \BibitemOpen
    \bibfield  {author} {\bibinfo {author} {\bibfnamefont {A.}~\bibnamefont
            {Auff{\'e}ves}}, \bibinfo {author} {\bibfnamefont {D.}~\bibnamefont
            {Gerace}}, \bibinfo {author} {\bibfnamefont {S.}~\bibnamefont {Portolan}},
        \bibinfo {author} {\bibfnamefont {A.}~\bibnamefont {Drezet}}, \ and\ \bibinfo
        {author} {\bibfnamefont {M.~F.}\ \bibnamefont {Santos}},\ }\href@noop {}
    {\bibfield  {journal} {\bibinfo  {journal} {New J. Phys.}\ }\textbf {\bibinfo
            {volume} {13}},\ \bibinfo {pages} {093020} (\bibinfo {year}
        {2011})}\BibitemShut {NoStop}%
    \bibitem [{\citenamefont {Radulaski}\ \emph {et~al.}(2017)\citenamefont
        {Radulaski}, \citenamefont {Fischer}, \citenamefont {Lagoudakis},
        \citenamefont {Zhang},\ and\ \citenamefont
        {Vu$\check{\rm{c}}$kovi$\grave{\rm{c}}$}}]{Radulaski2017}%
    \BibitemOpen
    \bibfield  {author} {\bibinfo {author} {\bibfnamefont {M.}~\bibnamefont
            {Radulaski}}, \bibinfo {author} {\bibfnamefont {K.~A.}\ \bibnamefont
            {Fischer}}, \bibinfo {author} {\bibfnamefont {K.~G.}\ \bibnamefont
            {Lagoudakis}}, \bibinfo {author} {\bibfnamefont {J.~L.}\ \bibnamefont
            {Zhang}}, \ and\ \bibinfo {author} {\bibfnamefont {J.}~\bibnamefont
            {Vu$\check{\rm{c}}$kovi$\grave{\rm{c}}$}},\ }\href@noop {} {\bibfield
        {journal} {\bibinfo  {journal} {Phys. Rev. A}\ }\textbf {\bibinfo {volume}
            {96}},\ \bibinfo {pages} {011801 (R)} (\bibinfo {year} {2017})}\BibitemShut
    {NoStop}%
    \bibitem [{\citenamefont {Meiser}\ and\ \citenamefont
        {Holland}(2010)}]{MeiserHolland2010a}%
    \BibitemOpen
    \bibfield  {author} {\bibinfo {author} {\bibfnamefont {D.}~\bibnamefont
            {Meiser}}\ and\ \bibinfo {author} {\bibfnamefont {M.~J.}\ \bibnamefont
            {Holland}},\ }\href@noop {} {\bibfield  {journal} {\bibinfo  {journal} {Phys.
                Rev. A}\ }\textbf {\bibinfo {volume} {81}},\ \bibinfo {pages} {063827}
        (\bibinfo {year} {2010})}\BibitemShut {NoStop}%
    \bibitem [{\citenamefont {Richter}\ \emph {et~al.}(2015)\citenamefont
        {Richter}, \citenamefont {Gegg}, \citenamefont {Theuerholz},\ and\
        \citenamefont {Knorr}}]{Richter2015}%
    \BibitemOpen
    \bibfield  {author} {\bibinfo {author} {\bibfnamefont {M.}~\bibnamefont
            {Richter}}, \bibinfo {author} {\bibfnamefont {M.}~\bibnamefont {Gegg}},
        \bibinfo {author} {\bibfnamefont {T.~S.}\ \bibnamefont {Theuerholz}}, \ and\
        \bibinfo {author} {\bibfnamefont {A.}~\bibnamefont {Knorr}},\ }\href@noop {}
    {\bibfield  {journal} {\bibinfo  {journal} {Phys. Rev. B}\ }\textbf {\bibinfo
            {volume} {91}},\ \bibinfo {pages} {035306} (\bibinfo {year}
        {2015})}\BibitemShut {NoStop}%
    \bibitem [{\citenamefont {Grankin}\ \emph {et~al.}(2014)\citenamefont
        {Grankin}, \citenamefont {Brion}, \citenamefont {Bimbard}, \citenamefont {Boddeda}, \citenamefont {Usmani}, \citenamefont {Ourjoumtsev},\ and\
        \citenamefont {Grangier}}]{Grangier2014}%
    \BibitemOpen
    \bibfield  {author} {\bibinfo {author} {\bibfnamefont {A.}~\bibnamefont
            {Grankin}}, \bibinfo {author} {\bibfnamefont {E.}~\bibnamefont {Brion}},
        \bibinfo {author} {\bibfnamefont {E.}\ \bibnamefont {Bimbard}}, \bibinfo {author} {\bibfnamefont {R.}\ \bibnamefont {Boddeda}}, \bibinfo {author} {\bibfnamefont {I.}\ \bibnamefont {Usmani}}, \bibinfo {author} {\bibfnamefont {A.}\ \bibnamefont {Ourjoumtsev}},\ and\
        \bibinfo {author} {\bibfnamefont {P.}~\bibnamefont {Grangier}},\ }\href@noop {}
    {\bibfield  {journal} {\bibinfo  {journal} {New J. Phys.}\ }\textbf {\bibinfo
            {volume} {16}},\ \bibinfo {pages} {043020} (\bibinfo {year}
        {2014})}\BibitemShut {NoStop}%
    \bibitem [{\citenamefont {Temnov}\ and\ \citenamefont
        {Woggon}(2005)}]{TemnovWoggon2005}%
    \BibitemOpen
    \bibfield  {author} {\bibinfo {author} {\bibfnamefont {V.~V.}\ \bibnamefont
            {Temnov}}\ and\ \bibinfo {author} {\bibfnamefont {U.}~\bibnamefont
            {Woggon}},\ }\href@noop {} {\bibfield  {journal} {\bibinfo  {journal} {Phys.
                Rev. Lett.}\ }\textbf {\bibinfo {volume} {95}},\ \bibinfo {pages} {243602}
        (\bibinfo {year} {2005})}\BibitemShut {NoStop}%
    \bibitem [{\citenamefont {Temnov}\ and\ \citenamefont
        {Woggon}(2009)}]{TemnovWoggon2009}%
    \BibitemOpen
    \bibfield  {author} {\bibinfo {author} {\bibfnamefont {V.~V.}\ \bibnamefont
            {Temnov}}\ and\ \bibinfo {author} {\bibfnamefont {U.}~\bibnamefont
            {Woggon}},\ }\href@noop {} {\bibfield  {journal} {\bibinfo  {journal} {Opt.
                Express}\ }\textbf {\bibinfo {volume} {17}},\ \bibinfo {pages} {5774}
        (\bibinfo {year} {2009})}\BibitemShut {NoStop}%
    \bibitem [{\citenamefont {Gegg}\ and\ \citenamefont
        {Richter}(2016)}]{Richter2016}%
    \BibitemOpen
    \bibfield  {author} {\bibinfo {author} {\bibfnamefont {M.}~\bibnamefont
            {Gegg}}\ and\ \bibinfo {author} {\bibfnamefont {M.}~\bibnamefont {Richter}},\
    }\href@noop {} {\bibfield  {journal} {\bibinfo  {journal} {New J. Phys.}\
        }\textbf {\bibinfo {volume} {18}},\ \bibinfo {pages} {043037} (\bibinfo
        {year} {2016})}\BibitemShut {NoStop}%
    \bibitem [{\citenamefont {Savasta}\ \emph {et~al.}(2010)\citenamefont
        {Savasta}, \citenamefont {Saija}, \citenamefont {Ridolfo}, \citenamefont
        {Stefano}, \citenamefont {Denti},\ and\ \citenamefont
        {Borghese}}]{SavastaACS2010}%
    \BibitemOpen
    \bibfield  {author} {\bibinfo {author} {\bibfnamefont {S.}~\bibnamefont
            {Savasta}}, \bibinfo {author} {\bibfnamefont {R.}~\bibnamefont {Saija}},
        \bibinfo {author} {\bibfnamefont {A.}~\bibnamefont {Ridolfo}}, \bibinfo
        {author} {\bibfnamefont {O.~D.}\ \bibnamefont {Stefano}}, \bibinfo {author}
        {\bibfnamefont {P.}~\bibnamefont {Denti}}, \ and\ \bibinfo {author}
        {\bibfnamefont {F.}~\bibnamefont {Borghese}},\ }\href@noop {} {\bibfield
        {journal} {\bibinfo  {journal} {ACS Nano}\ }\textbf {\bibinfo {volume} {4}},\
        \bibinfo {pages} {6369} (\bibinfo {year} {2010})}\BibitemShut {NoStop}%
    \bibitem [{\citenamefont {Chen}\ \emph {et~al.}(2013)\citenamefont {Chen},
        \citenamefont {Sandoghdar},\ and\ \citenamefont {Agio}}]{Sandoghdar2013}%
    \BibitemOpen
    \bibfield  {author} {\bibinfo {author} {\bibfnamefont {X.-W.}\ \bibnamefont
            {Chen}}, \bibinfo {author} {\bibfnamefont {V.}~\bibnamefont {Sandoghdar}}, \
        and\ \bibinfo {author} {\bibfnamefont {M.}~\bibnamefont {Agio}},\ }\href@noop
    {} {\bibfield  {journal} {\bibinfo  {journal} {Phys. Rev. Lett.}\ }\textbf
        {\bibinfo {volume} {110}},\ \bibinfo {pages} {153605} (\bibinfo {year}
        {2013})}\BibitemShut {NoStop}%
    \bibitem [{\citenamefont {Zengin}\ \emph {et~al.}(2015)\citenamefont {Zengin},
        \citenamefont {Wers{\"a}ll}, \citenamefont {Nilsson}, \citenamefont
        {Antosiewicz}, \citenamefont {K{\"a}ll},\ and\ \citenamefont
        {Shegai}}]{Shegai2015}%
    \BibitemOpen
    \bibfield  {author} {\bibinfo {author} {\bibfnamefont {G.}~\bibnamefont
            {Zengin}}, \bibinfo {author} {\bibfnamefont {M.}~\bibnamefont {Wers{\"a}ll}},
        \bibinfo {author} {\bibfnamefont {S.}~\bibnamefont {Nilsson}}, \bibinfo
        {author} {\bibfnamefont {T.~J.}\ \bibnamefont {Antosiewicz}}, \bibinfo
        {author} {\bibfnamefont {M.}~\bibnamefont {K{\"a}ll}}, \ and\ \bibinfo
        {author} {\bibfnamefont {T.}~\bibnamefont {Shegai}},\ }\href@noop {}
    {\bibfield  {journal} {\bibinfo  {journal} {Phys. Rev. Lett.}\ }\textbf
        {\bibinfo {volume} {114}},\ \bibinfo {pages} {157401} (\bibinfo {year}
        {2015})}\BibitemShut {NoStop}%
    \bibitem [{\citenamefont {Peyskens}\ \emph {et~al.}(2017)\citenamefont {Peyskens},
        \citenamefont {Chang}, \citenamefont {Englund}}]{Peyskens2017}%
    \BibitemOpen
    \bibfield  {author} {\bibinfo {author} {\bibfnamefont {F.}~\bibnamefont
            {Peyskens}}, \bibinfo {author} {\bibfnamefont {D.}~\bibnamefont {Chang}},\ and\
        \bibinfo {author} {\bibfnamefont {D.}~\bibnamefont {Englund}},\ }\href@noop {}
    {\bibfield  {journal} {\bibinfo  {journal} {Phys. Rev. B}\ }\textbf
        {\bibinfo {volume} {96}},\ \bibinfo {pages} {235151} (\bibinfo {year}
        {2017})}\BibitemShut {NoStop}%
    \bibitem [{\citenamefont {Tame}\ \emph {et~al.}(2013)\citenamefont {Tame},
        \citenamefont {McEnery}, \citenamefont {{\"O}zdemir}, \citenamefont {Lee},
        \citenamefont {Maier},\ and\ \citenamefont {Kim}}]{Tame2013}%
    \BibitemOpen
    \bibfield  {author} {\bibinfo {author} {\bibfnamefont {M.~S.}\ \bibnamefont
            {Tame}}, \bibinfo {author} {\bibfnamefont {K.~R.}\ \bibnamefont {McEnery}},
        \bibinfo {author} {\bibfnamefont {S.~K.}\ \bibnamefont {{\"O}zdemir}},
        \bibinfo {author} {\bibfnamefont {J.}~\bibnamefont {Lee}}, \bibinfo {author}
        {\bibfnamefont {S.~A.}\ \bibnamefont {Maier}}, \ and\ \bibinfo {author}
        {\bibfnamefont {M.~S.}\ \bibnamefont {Kim}},\ }\href@noop {} {\bibfield
        {journal} {\bibinfo  {journal} {Nature Physics}\ }\textbf {\bibinfo {volume}
            {9}},\ \bibinfo {pages} {329} (\bibinfo {year} {2013})}\BibitemShut {NoStop}%
    \bibitem [{\citenamefont {Akahane}\ \emph
        {et~al.}(2003{\natexlab{b}})\citenamefont {Akahane}, \citenamefont {Asano},
        \citenamefont {Song},\ and\ \citenamefont {Noda}}]{Noda2003}%
    \BibitemOpen
    \bibfield  {author} {\bibinfo {author} {\bibfnamefont {Y.}~\bibnamefont
            {Akahane}}, \bibinfo {author} {\bibfnamefont {T.}~\bibnamefont {Asano}},
        \bibinfo {author} {\bibfnamefont {B.-S.}\ \bibnamefont {Song}}, \ and\
        \bibinfo {author} {\bibfnamefont {S.}~\bibnamefont {Noda}},\ }\href@noop {}
    {\bibfield  {journal} {\bibinfo  {journal} {Nature}\ }\textbf {\bibinfo
            {volume} {425}},\ \bibinfo {pages} {944} (\bibinfo {year}
        {2003}{\natexlab{b}})}\BibitemShut {NoStop}%
    \bibitem [{\citenamefont {Englund}\ \emph {et~al.}(2007)\citenamefont
        {Englund}, \citenamefont {Faraon}, \citenamefont {Fushman}, \citenamefont
        {Stoltz}, \citenamefont {Petroff},\ and\ \citenamefont
        {Vu$\check{\rm{c}}$kovi$\grave{\rm{c}}$}}]{Englund2007}%
    \BibitemOpen
    \bibfield  {author} {\bibinfo {author} {\bibfnamefont {D.}~\bibnamefont
            {Englund}}, \bibinfo {author} {\bibfnamefont {A.}~\bibnamefont {Faraon}},
        \bibinfo {author} {\bibfnamefont {I.}~\bibnamefont {Fushman}}, \bibinfo
        {author} {\bibfnamefont {N.}~\bibnamefont {Stoltz}}, \bibinfo {author}
        {\bibfnamefont {P.}~\bibnamefont {Petroff}}, \ and\ \bibinfo {author}
        {\bibfnamefont {J.}~\bibnamefont {Vu$\check{\rm{c}}$kovi$\grave{\rm{c}}$}},\
    }\href@noop {} {\bibfield  {journal} {\bibinfo  {journal} {Nature}\ }\textbf
        {\bibinfo {volume} {450}},\ \bibinfo {pages} {857} (\bibinfo {year}
        {2007})}\BibitemShut {NoStop}%
    \bibitem [{\citenamefont {Jewell}\ \emph {et~al.}(1989)\citenamefont {Jewell},
        \citenamefont {McCall}, \citenamefont {Lee}, \citenamefont {Scherer},
        \citenamefont {Gossard},\ and\ \citenamefont {English}}]{Jewell1989}%
    \BibitemOpen
    \bibfield  {author} {\bibinfo {author} {\bibfnamefont {J.~L.}\ \bibnamefont
            {Jewell}}, \bibinfo {author} {\bibfnamefont {S.~L.}\ \bibnamefont {McCall}},
        \bibinfo {author} {\bibfnamefont {Y.~H.}\ \bibnamefont {Lee}}, \bibinfo
        {author} {\bibfnamefont {A.}~\bibnamefont {Scherer}}, \bibinfo {author}
        {\bibfnamefont {A.~C.}\ \bibnamefont {Gossard}}, \ and\ \bibinfo {author}
        {\bibfnamefont {J.~H.}\ \bibnamefont {English}},\ }\href@noop {} {\bibfield
        {journal} {\bibinfo  {journal} {Appl. Phys. Lett.}\ }\textbf {\bibinfo
            {volume} {54}},\ \bibinfo {pages} {1400} (\bibinfo {year}
        {1989})}\BibitemShut {NoStop}%
    \bibitem [{\citenamefont {Walls}\ and\ \citenamefont
        {Milburn}(2008{\natexlab{a}})}]{WallsMilburnBook}%
    \BibitemOpen
    \bibfield  {author} {\bibinfo {author} {\bibfnamefont {D.~F.}\ \bibnamefont
            {Walls}}\ and\ \bibinfo {author} {\bibfnamefont {G.~J.}\ \bibnamefont
            {Milburn}},\ }\href@noop {} {\emph {\bibinfo {title} {Quantum Optics}}},\
    \bibinfo {edition} {2nd}\ ed.\ (\bibinfo  {publisher} {Springer},\ \bibinfo
    {address} {Berlin},\ \bibinfo {year} {2008})\ \bibinfo {note} {p
        199}\BibitemShut {NoStop}%
    \bibitem [{\citenamefont {Loudon}(1973)}]{LoudonBook}%
    \BibitemOpen
    \bibfield  {author} {\bibinfo {author} {\bibfnamefont {R.}~\bibnamefont
            {Loudon}},\ }\href@noop {} {\emph {\bibinfo {title} {The quantum theory of
                light}}}\ (\bibinfo  {publisher} {Oxford University Press},\ \bibinfo
    {address} {Oxford},\ \bibinfo {year} {1973})\BibitemShut {NoStop}%
    \bibitem [{\citenamefont {Zou}\ and\ \citenamefont
        {Mandel}(1990)}]{ZouMandel1990}%
    \BibitemOpen
    \bibfield  {author} {\bibinfo {author} {\bibfnamefont {X.~T.}\ \bibnamefont
            {Zou}}\ and\ \bibinfo {author} {\bibfnamefont {L.}~\bibnamefont {Mandel}},\
    }\href@noop {} {\bibfield  {journal} {\bibinfo  {journal} {Phys. Rev. A}\
        }\textbf {\bibinfo {volume} {41}},\ \bibinfo {pages} {475} (\bibinfo {year}
        {1990})}\BibitemShut {NoStop}%
    \bibitem [{\citenamefont {Walls}\ and\ \citenamefont
        {Milburn}(2008{\natexlab{b}})}]{WallsMilburnBookSTATISTICS}%
    \BibitemOpen
    \bibfield  {author} {\bibinfo {author} {\bibfnamefont {D.~F.}\ \bibnamefont
            {Walls}}\ and\ \bibinfo {author} {\bibfnamefont {G.~J.}\ \bibnamefont
            {Milburn}},\ }\href@noop {} {\emph {\bibinfo {title} {Quantum Optics}}},\
    \bibinfo {edition} {2nd}\ ed.\ (\bibinfo  {publisher} {Springer},\ \bibinfo
    {address} {Berlin},\ \bibinfo {year} {2008})\ \bibinfo {note} {pp
        38-42}\BibitemShut {NoStop}%
    \bibitem [{\citenamefont {Novotny}\ and\ \citenamefont
        {Hecht}(2006)}]{NovotnyBook}%
    \BibitemOpen
    \bibfield  {author} {\bibinfo {author} {\bibfnamefont {L.}~\bibnamefont
            {Novotny}}\ and\ \bibinfo {author} {\bibfnamefont {B.}~\bibnamefont
            {Hecht}},\ }\href@noop {} {\emph {\bibinfo {title} {Principles of
                nano-optics}}}\ (\bibinfo  {publisher} {Cambridge University Press},\
    \bibinfo {address} {Cambridge, England},\ \bibinfo {year} {2006})\ \bibinfo
    {note} {p 276}\BibitemShut {NoStop}%
    \bibitem [{\citenamefont {Maier}(2007)}]{MaierBook}%
    \BibitemOpen
    \bibfield  {author} {\bibinfo {author} {\bibfnamefont {S.~A.}\ \bibnamefont
            {Maier}},\ }\href@noop {} {\emph {\bibinfo {title} {Plasmonics: Fundamentals
                and applications}}}\ (\bibinfo  {publisher} {Springer},\ \bibinfo {address}
    {New York},\ \bibinfo {year} {2007})\ \bibinfo {note} {pp 73-77}\BibitemShut
    {NoStop}%
    \bibitem [{\citenamefont {Gonz{\'a}lez-Ballestero}\ \emph
        {et~al.}(2015)\citenamefont {Gonz{\'a}lez-Ballestero}, \citenamefont {Feist},
        \citenamefont {Moreno},\ and\ \citenamefont
        {Garc{\'i}a-Vidal}}]{GlezBallestero2015}%
    \BibitemOpen
    \bibfield  {author} {\bibinfo {author} {\bibfnamefont {C.}~\bibnamefont
            {Gonz{\'a}lez-Ballestero}}, \bibinfo {author} {\bibfnamefont
            {J.}~\bibnamefont {Feist}}, \bibinfo {author} {\bibfnamefont
            {E.}~\bibnamefont {Moreno}}, \ and\ \bibinfo {author} {\bibfnamefont {F.~J.}\
            \bibnamefont {Garc{\'i}a-Vidal}},\ }\href@noop {} {\bibfield  {journal}
        {\bibinfo  {journal} {Phys. Rev. B}\ }\textbf {\bibinfo {volume} {92}},\
        \bibinfo {pages} {121402(R)} (\bibinfo {year} {2015})}\BibitemShut {NoStop}%
    \bibitem [{\citenamefont {Thoma}\ \emph {et~al.}(2016)\citenamefont {Thoma},
        \citenamefont {Schnauber}, \citenamefont {Gschrey}, \citenamefont {Seifried},
        \citenamefont {Wolters}, \citenamefont {Schulze}, \citenamefont
        {Strittmatter}, \citenamefont {Rodt}, \citenamefont {Carmele}, \citenamefont
        {Knorr}, \citenamefont {Heindel},\ and\ \citenamefont
        {Reitzenstein}}]{Thoma2016}%
    \BibitemOpen
    \bibfield  {author} {\bibinfo {author} {\bibfnamefont {A.}~\bibnamefont
            {Thoma}}, \bibinfo {author} {\bibfnamefont {P.}~\bibnamefont {Schnauber}},
        \bibinfo {author} {\bibfnamefont {M.}~\bibnamefont {Gschrey}}, \bibinfo
        {author} {\bibfnamefont {M.}~\bibnamefont {Seifried}}, \bibinfo {author}
        {\bibfnamefont {J.}~\bibnamefont {Wolters}}, \bibinfo {author} {\bibfnamefont
            {J.-H.}\ \bibnamefont {Schulze}}, \bibinfo {author} {\bibfnamefont
            {A.}~\bibnamefont {Strittmatter}}, \bibinfo {author} {\bibfnamefont
            {S.}~\bibnamefont {Rodt}}, \bibinfo {author} {\bibfnamefont {A.}~\bibnamefont
            {Carmele}}, \bibinfo {author} {\bibfnamefont {A.}~\bibnamefont {Knorr}},
        \bibinfo {author} {\bibfnamefont {T.}~\bibnamefont {Heindel}}, \ and\
        \bibinfo {author} {\bibfnamefont {S.}~\bibnamefont {Reitzenstein}},\
    }\href@noop {} {\bibfield  {journal} {\bibinfo  {journal} {Phys. Rev. Lett.}\
        }\textbf {\bibinfo {volume} {116}},\ \bibinfo {pages} {033601} (\bibinfo
        {year} {2016})}\BibitemShut {NoStop}%
    \bibitem [{\citenamefont {Rector}\ and\ \citenamefont
        {Faye}(1998)}]{Rector1998}%
    \BibitemOpen
    \bibfield  {author} {\bibinfo {author} {\bibfnamefont {K.~D.}\ \bibnamefont
            {Rector}}\ and\ \bibinfo {author} {\bibfnamefont {M.~D.}\ \bibnamefont
            {Faye}},\ }\href@noop {} {\bibfield  {journal} {\bibinfo  {journal} {J. Chem.
                Phys.}\ }\textbf {\bibinfo {volume} {108}},\ \bibinfo {pages} {1794}
        (\bibinfo {year} {1998})}\BibitemShut {NoStop}%
    \bibitem [{\citenamefont {Zhang}\ \emph {et~al.}(2006)\citenamefont {Zhang},
        \citenamefont {Govorov},\ and\ \citenamefont {Bryant}}]{Bryant2006}%
    \BibitemOpen
    \bibfield  {author} {\bibinfo {author} {\bibfnamefont {W.}~\bibnamefont
            {Zhang}}, \bibinfo {author} {\bibfnamefont {A.~O.}\ \bibnamefont {Govorov}},
        \ and\ \bibinfo {author} {\bibfnamefont {G.~W.}\ \bibnamefont {Bryant}},\
    }\href@noop {} {\bibfield  {journal} {\bibinfo  {journal} {Phys. Rev. Lett.}\
        }\textbf {\bibinfo {volume} {97}},\ \bibinfo {pages} {146804} (\bibinfo
        {year} {2006})}\BibitemShut {NoStop}%
    \bibitem [{\citenamefont {Artuso}\ and\ \citenamefont
        {Bryant}(2008)}]{Bryant2008}%
    \BibitemOpen
    \bibfield  {author} {\bibinfo {author} {\bibfnamefont {R.~D.}\ \bibnamefont
            {Artuso}}\ and\ \bibinfo {author} {\bibfnamefont {G.~W.}\ \bibnamefont
            {Bryant}},\ }\href@noop {} {\bibfield  {journal} {\bibinfo  {journal} {Nano
                Letters}\ }\textbf {\bibinfo {volume} {8}},\ \bibinfo {pages} {2106}
        (\bibinfo {year} {2008})}\BibitemShut {NoStop}%
    \bibitem [{\citenamefont {Ridolfo}\ \emph {et~al.}(2010)\citenamefont
        {Ridolfo}, \citenamefont {Di~Stefano}, \citenamefont {Fina}, \citenamefont
        {Saija},\ and\ \citenamefont {Savasta}}]{SavastaPRL2010}%
    \BibitemOpen
    \bibfield  {author} {\bibinfo {author} {\bibfnamefont {A.}~\bibnamefont
            {Ridolfo}}, \bibinfo {author} {\bibfnamefont {O.}~\bibnamefont {Di~Stefano}},
        \bibinfo {author} {\bibfnamefont {N.}~\bibnamefont {Fina}}, \bibinfo {author}
        {\bibfnamefont {R.}~\bibnamefont {Saija}}, \ and\ \bibinfo {author}
        {\bibfnamefont {S.}~\bibnamefont {Savasta}},\ }\href@noop {} {\bibfield
        {journal} {\bibinfo  {journal} {Phys. Rev. Lett.}\ }\textbf {\bibinfo
            {volume} {105}} (\bibinfo {year} {2010})}\BibitemShut {NoStop}%
    \bibitem [{\citenamefont {Liew}\ and\ \citenamefont
        {Savona}(2010)}]{Savona2010}%
    \BibitemOpen
    \bibfield  {author} {\bibinfo {author} {\bibfnamefont {T.~C.~H.}\
            \bibnamefont {Liew}}\ and\ \bibinfo {author} {\bibfnamefont {V.}~\bibnamefont
            {Savona}},\ }\href@noop {} {\bibfield  {journal} {\bibinfo  {journal} {Phys.
                Rev. Lett.}\ }\textbf {\bibinfo {volume} {104}} (\bibinfo {year}
        {2010})}\BibitemShut {NoStop}%
    \bibitem [{\citenamefont {Bamba}\ \emph {et~al.}(2011)\citenamefont {Bamba},
        \citenamefont {Imamo$\breve{\g}$lu}, \citenamefont {Carusotto},\ and\
        \citenamefont {Ciuti}}]{Ciuti2011}%
    \BibitemOpen
    \bibfield  {author} {\bibinfo {author} {\bibfnamefont {M.}~\bibnamefont
            {Bamba}}, \bibinfo {author} {\bibfnamefont {A.}~\bibnamefont
            {Imamo$\breve{\g}$lu}}, \bibinfo {author} {\bibfnamefont {I.}~\bibnamefont
            {Carusotto}}, \ and\ \bibinfo {author} {\bibfnamefont {C.}~\bibnamefont
            {Ciuti}},\ }\href@noop {} {\bibfield  {journal} {\bibinfo  {journal} {Phys.
                Rev. A}\ }\textbf {\bibinfo {volume} {83}},\ \bibinfo {pages} {021802(R)}
        (\bibinfo {year} {2011})}\BibitemShut {NoStop}%
    \bibitem [{\citenamefont {Carusotto}\ and\ \citenamefont
        {Ciuti}(2013)}]{Ciuti2013}%
    \BibitemOpen
    \bibfield  {author} {\bibinfo {author} {\bibfnamefont {I.}~\bibnamefont
            {Carusotto}}\ and\ \bibinfo {author} {\bibfnamefont {C.}~\bibnamefont
            {Ciuti}},\ }\href@noop {} {\bibfield  {journal} {\bibinfo  {journal} {Rev.
                Mod. Phys.}\ }\textbf {\bibinfo {volume} {85}},\ \bibinfo {pages} {299}
        (\bibinfo {year} {2013})}\BibitemShut {NoStop}%
\end{thebibliography}

%

\end{document}